\documentclass{IEEEtran}
\usepackage{cite}
\ifCLASSINFOpdf
   \usepackage[pdftex]{graphicx}
   \DeclareGraphicsExtensions{.pdf,.jpeg,.png}
\else
   \usepackage[dvips]{graphicx}
   \graphicspath{{./figures/}}
   \DeclareGraphicsExtensions{.eps}
\fi
\usepackage[cmex10]{amsmath}
\usepackage{amssymb}
\usepackage{bm}
\usepackage{subfigure}
\usepackage{mathtools}

\newtheorem{theorem}{Theorem}

\newtheorem{proposition}{Proposition}
\newtheorem{lemma}{Lemma}
\newtheorem{corollary}{Corollary}
\newtheorem{definition}{Definition}
\newtheorem{remark}{Remark}
\newtheorem{example}{Example}

\def\sumnetalg{SUM-NET-CONS~}
\newcommand{\beqno}{ \begin{equation*} }
\newcommand{\eeqno}{ \end{equation*} }
\newcommand{\beq}{ \begin{equation} }
\newcommand{\eeq}{ \end{equation} }

\newcommand{\calB}{\mathcal{B}}
\newcommand{\calP}{\mathcal{P}}
\newcommand{\calD}{\mathcal{D}}
\newcommand{\calI}{\mathcal{I}}

\DeclareMathOperator{\ch}{ch}
\DeclareMathOperator{\diag}{diag}
\DeclareMathOperator{\In}{In}
\DeclareMathOperator{\tail}{tail}
\DeclareMathOperator{\head}{head}
\DeclareMathOperator{\rank}{rank}
\DeclareMathOperator{\capacity}{cap}
\DeclareMathOperator{\Span}{Span}
\usepackage{algorithm}
\usepackage{algorithmic}

\usepackage{array}
\begin{document}
%
% paper title
% Titles are generally capitalized except for words such as a, an, and, as,
% at, but, by, for, in, nor, of, on, or, the, to and up, which are usually
% not capitalized unless they are the first or last word of the title.
% Linebreaks \\ can be used within to get better formatting as desired.
% Do not put math or special symbols in the title.
\title{Sum-networks from incidence structures: construction and capacity analysis}
%
%
% author names and IEEE memberships
% note positions of commas and nonbreaking spaces ( ~ ) LaTeX will not break
% a structure at a ~ so this keeps an author's name from being broken across
% two lines.
% use \thanks{} to gain access to the first footnote area
% a separate \thanks must be used for each paragraph as LaTeX2e's \thanks
% was not built to handle multiple paragraphs
%

\author{Ardhendu~Tripathy,~\IEEEmembership{Student Member,~IEEE,}
%        John~Doe,~\IEEEmembership{Fellow,~OSA,}
        and~Aditya~Ramamoorthy,~\IEEEmembership{Member,~IEEE}% <-this % stops a space
\thanks{ This work was supported in part by the
National Science Foundation (NSF) under grants CCF-1149860,
CCF-1320416 and CCF-1718470. The material in this work has appeared in part at the 52nd Allerton Conference on Communication, Control and Computing, 2014 and the 2015 IEEE International Symposium on Information Theory. The authors are with the Department
of Electrical and Computer Engineering, Iowa State University, Ames,
IA, 50011 USA. e-mail: \{ardhendu@iastate.edu, adityar@iastate.edu\}
}
}

% note the % following the last \IEEEmembership and also \thanks -
% these prevent an unwanted space from occurring between the last author name
% and the end of the author line. i.e., if you had this:
%
% \author{....lastname \thanks{...} \thanks{...} }
%                     ^------------^------------^----Do not want these spaces!
%
% a space would be appended to the last name and could cause every name on that
% line to be shifted left slightly. This is one of those "LaTeX things". For
% instance, "\textbf{A} \textbf{B}" will typeset as "A B" not "AB". To get
% "AB" then you have to do: "\textbf{A}\textbf{B}"
% \thanks is no different in this regard, so shield the last } of each \thanks
% that ends a line with a % and do not let a space in before the next \thanks.
% Spaces after \IEEEmembership other than the last one are OK (and needed) as
% you are supposed to have spaces between the names. For what it is worth,
% this is a minor point as most people would not even notice if the said evil
% space somehow managed to creep in.

% The paper headers
\markboth{IEEE Transactions on Information Theory}%
{Tripathy \MakeLowercase{\textit{et al.}}: On the zero-error computation capacity of sum-networks obtained using incidence structures}
% The only time the second header will appear is for the odd numbered pages
% after the title page when using the twoside option.
%
% *** Note that you probably will NOT want to include the author's ***
% *** name in the headers of peer review papers.                   ***
% You can use \ifCLASSOPTIONpeerreview for conditional compilation here if
% you desire.

% If you want to put a publisher's ID mark on the page you can do it like
% this:
%\IEEEpubid{0000--0000/00\$00.00~\copyright~2014 IEEE}
% Remember, if you use this you must call \IEEEpubidadjcol in the second
% column for its text to clear the IEEEpubid mark.

% use for special paper notices
%\IEEEspecialpapernotice{(Invited Paper)}

% make the title area
\maketitle

% As a general rule, do not put math, special symbols or citations
% in the abstract or keywords.
\begin{abstract}
A sum-network is an instance of a function computation problem %network coding problem
over a directed acyclic network in which each terminal node wants to compute the sum over a finite field of the information observed at all the source nodes. Many characteristics of the well-studied multiple unicast network communication problem also hold for sum-networks due to a known reduction between the %instances of these
two problems. In this work, we describe an algorithm to construct families of sum-network instances using \textit{incidence structures}. The computation capacity of several of these sum-network families is evaluated. %characterized. %We demonstrate that
Unlike the coding capacity of a multiple unicast problem, the computation capacity of sum-networks depends on the characteristic of the finite field over which the sum is computed. This dependence is very strong; we show examples of sum-networks that have a rate-1 solution over one characteristic but a rate close to zero over a different characteristic. Additionally, a sum-network can have arbitrarily different computation capacities for different alphabets.% This is in contrast to the multiple unicast problem where it is known that the capacity is independent of the network coding alphabet.
\end{abstract}

% Note that keywords are not normally used for peerreview papers.
\begin{IEEEkeywords}
network coding, function computation, sum-networks, characteristic, incidence structures
\end{IEEEkeywords}

% For peer review papers, you can put extra information on the cover
% page as needed:
% \ifCLASSOPTIONpeerreview
% \begin{center} \bfseries EDICS Category: 3-BBND \end{center}
% \fi
%
% For peerreview papers, this IEEEtran command inserts a page break and
% creates the second title. It will be ignored for other modes.
\IEEEpeerreviewmaketitle

\section{Introduction}
Applications as diverse as parallel processing, distributed data analytics and sensor networks often deal with variants of the problem of distributed computation. This has motivated the study of various problems in the fields of computer science, automatic control and information theory. Broadly speaking, one can model this question in the following manner. Consider a directed acyclic network with its edges denoting communication links. A subset of the nodes observe certain information, these nodes are called sources. A different subset of nodes, called terminals, wish to compute functions of the observed information with a certain fidelity. The computation is carried out by the terminals with the aid of the information received over their incoming edges. The demand functions and the network topology are a part of the problem instance and can be arbitrary. This framework is very general and encompasses several problems that have received significant research attention.

Prior work \cite{kornerM79},\cite{orlitskyR01},\cite{doshiSME10} concerning information theoretic issues in function computation worked under the setting of correlated information observed at the sources and \textit{simple} network structures, which were simple in the sense that there were edges connecting the sources to the terminal without any intermediate nodes or relays. For instance, \cite{orlitskyR01} characterizes the amount of information that a source must transmit so that a terminal with some correlated side-information can reliably compute a function of the message observed at the source and the side-information. Reference \cite{doshiSME10} considered distributed functional compression, in which two messages are separately encoded and given to a decoder that computes a function of the two messages with an arbitrarily small probability of error.

With the advent of network coding \cite{al},\cite{koetterM03}, the scope of the questions considered included the setting in which the information observed at the sources is independent and the network topology is more complex. Under this setting, information is sent from a source to a terminal over a path of edges in the directed acyclic network with one or more intermediate nodes in it, these relay nodes have no limit on their memory or computational power. The communication edges are abstracted into error-free, delay-free links with a certain capacity for information transfer and are sometimes referred to as \textit{bit-pipes}. The messages are required to be recovered with zero distortion.
The \textit{multicast} scenario, in which the message observed at the only source in the network is demanded by all terminals in the network, is solved in \cite{al},\cite{koetterM03},\cite{liYC03}. A sufficient condition for solvability in the multicast scenario is that each terminal has a max-flow from the source that is at least the entropy rate of the message random process \cite{al}. Reference \cite{liYC03} established that \textit{linear} network codes over a sufficiently large alphabet can solve this problem and \cite{koetterM03} provided necessary and sufficient conditions for solving a multicast problem instance in an algebraic framework. The work in \cite{koetterM03} also gave a simple algorithm to construct a network code that satisfies it.

Unlike the multicast problem, the multiple unicast problem does not admit such a clean solution.
This scenario has multiple source-terminal pairs over a directed acyclic network of bit-pipes and each terminal wants to recover the message sent by its corresponding source with the help of the information transmitted on the network.
To begin with, there are problem instances where more than one use of the network is required to solve it. To model this, each network edge is viewed as carrying a vector of $n$ alphabet symbols, while each message is a vector of $m$ alphabet symbols. A network code specifies the relationship between the vector transmitted on each edge of the network and the message vectors, and it solves a network coding problem instance if $m=n$. It is shown that linear network codes are in general not sufficient to solve this problem \cite{doughertyFZ05}. One can define the notion of \textit{coding capacity} of a network as the supremum of the ratio $m/n$ over all network codes that allow each terminal to recover its desired message; this ratio $m/n$ for a particular network code is called its \textit{rate}. The coding capacity of a network is independent of the alphabet used \cite{CannonsDFZ06}. While a network code with any rational rate less than the coding capacity exists by definition and zero-padding, a network code with rate equal to coding capacity does not exist for certain networks, even if the coding capacity is rational \cite{doughertyFZ06}. The multi-commodity flow solution to the multiple unicast problem is called a routing solution, as the different messages can be interpreted as distinct commodities \textit{routed} through the intermediate nodes. It is well-known that in the case of multicast, network coding can provide a gain in rate over traditional routing of messages that scales with the size of the network \cite{jaggiSCEEJT05}. However, evaluating the coding capacity for an arbitrary instance of the network coding problem is known to be hard in general \cite{huangR12_TCOM}, \cite{huangR14}, \cite{lehmanL04}, \cite{kamathTW14}.

Expanding the scope of the demands of the terminals, \cite{appuswamyFKZ11} considered \textit{function computation} over directed acyclic networks with only one terminal; the value to be recovered at the terminal was allowed to be a function of the messages as opposed to being a subset of the set of all messages. This computation is performed using information transmitted over the edges by a network code. Analogous to the coding capacity, a notion of \textit{computation capacity} can be defined in this case. A rate-$m/n$ network code that allows the terminal to compute its demand function has the interpretation that the function can be computed by the terminal $m$ times in $n$ uses of the network. Cut-set based upper bounds for the computation capacity of a directed acyclic network with one terminal were given in \cite{appuswamyFKZ11},\cite{huangTY15}. A matching lower bound for function computation in tree-networks was given in \cite{appuswamyFKZ11} and the computation capacity of linear and non-linear network codes for different \textit{classes} of demand functions was explored in \cite{appuswamyFKZ13}.

A different flavor of the function computation problem, often called the \textit{sum-network} problem, considers directed acyclic networks with multiple terminals, each of which demands the finite-field sum of all the messages observed at the sources \cite{ramamoorthy08}, \cite{raiD12}. Reference\cite{ramamoorthyL13} characterized the requirements that sum-networks with two or three sources or terminals must satisfy so that each terminal can recover the sum at unit rate. Similar to the network coding scenario, a sum-network whose terminals are satisfied by a rate-1 network code are called solvable sum-networks. Reference \cite{raiD12} established that deciding whether an arbitrary instance of a sum-network problem instance is solvable is at least as hard as deciding whether a suitably defined multiple unicast instance is solvable. As a result of this reduction the various characteristics of the solvability problem for network coding instances are also true for the solvability problem for sum-networks; this establishes the broadness of the class of sum-networks within all communication problems on directed acyclic networks.

While solvable sum-networks and solvable network coding instances are intimately related, the results in this paper indicate that these classes of problems diverge when we focus on coding/computation capacity, which can be strictly less than one. In \cite[Section VI]{CannonsDFZ06}, the coding capacity of networks is shown to be independent of the finite field chosen as the alphabet for the messages and the information transmitted over the edges. We show that an analogous statement is not true for sum-networks by demonstrating infinite families of sum-network problem instances whose computation capacity vary depending on the finite field alphabet. Moreover, the gap in computation capacity on two different finite fields is shown to scale with the network size for certain classes of sum-networks. For two alphabets $\mathcal{F}_1, \mathcal{F}_2$ of different cardinality and a network $\mathcal{N}$, the authors in \cite[Theorem VI.5]{CannonsDFZ06} described a procedure to simulate a rate-$m_2/n_2$ network code on $\mathcal{F}_2$ for $\mathcal{N}$ using a rate-$m_1/n_1$ network code on $\mathcal{F}_1$ for the same network, such that $m_2/n_2 \geq (m_1/n_1)-\epsilon$ for any $\epsilon > 0$. That procedure does not apply for sum-networks. Along the lines of the counterexample given in \cite{ramamoorthyL13} regarding minimum max-flow connectivity required for solvability of sum-networks with three sources and terminals, we provide an infinite family of counterexamples that mandate certain value of max-flow connectivity to allow solvability (over some finite field) of a general sum-network with more than three sources and terminals. These sum-network problem instances are arrived at using a systematic construction procedure on combinatorial objects called \textit{incidence structures}. Incidence structures are structured set systems and include, e.g., graphs and combinatorial designs \cite{Stinson}. We note here that combinatorial designs have recently been used to address issues such as the construction of distributed storage systems \cite{OlmezR16,RR} and coded caching systems \cite{TangR16,TangR16_resolv,TangR17}.

This paper is organized as follows. Section \ref{sec:related_work} describes previous work related to the problem considered and summarizes the contributions. Section \ref{sec:setup} describes the problem model formally and Section \ref{sec:constr} describes the construction procedure we use to obtain the sum-network problem instances considered in this work. Section \ref{sec:upper_bd} gives an upper bound on the computation capacity of these sum-networks and Section \ref{sec:lower_bd} describes a method to obtain linear network codes that achieve the upper bound on rate for several families of the sum-networks constructed. Section \ref{sec:discussion} interprets the results in this paper and outlines the key conclusions drawn in this paper. Section \ref{sec:conclusion} concludes the paper and discusses avenues for future work.

\section{Background, Related Work and Summary of Contributions} \label{sec:related_work}
The problem setting in which we will work is such that the information observed at the sources are independent and uniformly distributed over a finite field alphabet $\mathcal{F}$. The network links are error-free and assumed to have unit-capacity. Each of the possibly many terminals wants to recover the finite field sum of all the messages with zero error. This problem was introduced in the work of \cite{ramamoorthy08}. Intuitively, it is reasonable to assume the network resources, i.e., the capacity of the network links and the network structure have an effect on whether the sum can be computed successfully by all the terminals in the network. Reference \cite{ramamoorthyL13} characterized this notion for the class of sum-networks that have either two sources and/or two terminals. For this class of sum-networks it was shown that if the source messages had unit-entropy, a max-flow of one between each source-terminal pair was enough to solve the problem. It was shown by means of a counterexample that a max-flow of one was not enough to solve a sum-network with three sources and terminals. However, it was also shown that a max-flow of two between each source-terminal pair was sufficient to solve any sum-network with three sources and three terminals.
Reference \cite{rai13capacity} considered the computation capacity of the class of sum-networks that have three sources and three or more terminals or vice versa. It was shown that for any integer $k \geq 2$, there exist three-source, $n$-terminal sum-networks (where $n\geq 3$) whose computation capacity is $\frac{k}{k+1}$.
The work most closely related to this paper is \cite{raiD13}, which gives a construction procedure that for any positive rational number $p/q$ returns a sum-network whose computation capacity is $p/q$. Assuming that $p$ and $q$ are relatively prime, the procedure described in \cite{raiD13} constructs a sum-network that has $2q-1 + \binom{2q-1}{2}$ sources and $2q + \binom{2q-1}{2}$ terminals, which can be very large when $q$ is large. The authors asked the question if there exist smaller sum-networks (i.e., with fewer sources and terminals) that have the computation capacity as $p/q$. Our work in \cite{tripathyR14} answered it in the affirmative and proposed a general construction procedure that returned sum-networks with a prescribed computation capacity. The sum-networks in \cite{raiD13} could be obtained as special cases of this construction procedure. Some smaller instances of sum-networks for specific values were presented in \cite{dasR15}. Small sum-network instances can be useful in determining sufficiency conditions for larger networks.
The scope of the construction procedure proposed in \cite{tripathyR14} was widened in \cite{tripathyR15}, as a result of which, it was shown that there exist sum-network instances whose computation capacity depends rather strongly on the finite field alphabet.
This work builds on the contributions in \cite{tripathyR14,tripathyR15}. In particular, we present a systematic algebraic technique that encompasses the prior results. We also include proofs of all results and discuss the implications of our results in depth.
\subsection{Summary of contributions}
%\aditya{needs substantial rewriting}
In this work, we define several classes of sum-networks for which we can explicitly determine the computation capacity. These networks are constructed by using appropriately defined incidence structures. The main contributions of our work are as follows.
\begin{itemize}
\item We demonstrate novel techniques for determining upper and lower bounds on the computation capacity of the constructed sum-networks. In most cases, these bounds match, thus resulting in a determination of the capacity of these sum-networks.
\item We demonstrate a strong dependence of the computation capacity on the characteristic of the finite field over which the computation is taking place. In particular, for the {\it same} network, the computation capacity changes based on the characteristic of the underlying field. This is unlike the coding capacity for the multiple unicast problem which is known to be independent of the network coding alphabet.
\item Consider the class of networks where every source-terminal pair has a minimum cut of value at least $\alpha$, where $\alpha$ is an arbitrary positive integer. We demonstrate that there exists a sum-network within this class (with a large number of sources and terminals) whose computation capacity can be made arbitrarily small. This implies that the capacity of sum-networks cannot be characterized just by individual source-terminal minimum cuts.
\end{itemize}

\section{Problem formulation and Preliminaries}
\label{sec:setup}
We consider communication over a directed acyclic graph (DAG) $G=(V,E)$ where $V$ is the  set of nodes and $E \subseteq V \times V \times \mathbb{Z}_+$ are the edges denoting the delay-free communication links between them. The edges are given an additional index as the model allows for multiple edges between two distinct nodes. For instance, if there are two edges between nodes $u$ and $v$, these will be represented as $(u,v,1)$ and $(u,v,2)$. Subset $S \subset V$ denotes the source nodes and $T \subset V$ denotes the terminal nodes. The source nodes have no incoming edges and the terminal nodes have no outgoing edges. Each source node $s_i \in S$ observes an independent random process $X_i$, such that the sequence of random variables $X_{i1}, X_{i2}, \dots$ indexed by time (denoted by a positive integer) are i.i.d. and each $X_{ij}$ takes values that are uniformly distributed over a finite alphabet $\mathcal{F}$. The alphabet $\mathcal{F}$ is assumed to be a finite field with $|\mathcal{F}| = q$ and its characteristic denoted as $\ch(\mathcal{F})$. Each edge represents a communication channel of unit capacity, i.e., it can transmit one symbol from $\mathcal{F}$ per time slot. When referring to a communication link (or edge) without its third index, we will assume that it is the set of all edges between its two nodes. For such a set denoted by $(u,v)$, we define its capacity $\capacity(u,v)$ as the number of edges between $u$ and $v$.
We use the notation $\In(v)$ and $\In(e)$ to represent the set of incoming edges at node $v \in V$ and edge $e \in E$. For the edge $e=(u,v)$ let $\head(e)=v$ and $\tail(e)=u$. Each terminal node $t \in T$ demands the sum (over $\mathcal{F}$) of the individual source messages. Let $Z_j = \sum_{\{i: s_i \in S\}} X_{ij}$ for all $j \in \mathbb{N}$ (the set of natural numbers); then each $t \in T$ wants to recover the sequence $Z\coloneqq(Z_1,Z_2,\dots)$ from the information it receives on its incoming edges, i.e., the set $\In(t)$.

A network code is an assignment of local encoding functions to each edge $e \in E$ (denoted as $\tilde{\phi}_e({\cdot})$) and a decoding function to each terminal $t \in T$ (denoted as $\psi_t({\cdot})$) such that all the terminals can compute $Z$. The local encoding function for an edge connected to a set of sources only has the messages observed at those particular source nodes as its input arguments. Likewise, the input arguments for the local encoding function of an edge that is not connected to any source are the values received on its incoming edges and the inputs for the decoding function of a terminal are the values received on its incoming edges. As we consider directed acyclic networks, it can be seen that there is a \textit{global} encoding function that expresses the value transmitted on an edge in terms of the source messages in the set $X\coloneqq\{X_i: s_i \in S\}$. The global encoding function for an edge $e$ is denoted as $\phi_e(X)$.

The following notation describes the domain and range of the local encoding and decoding functions using two natural numbers $m$ and $n$ for a general vector network code. $m$ is the number of i.i.d. source values that are encoded simultaneously by the local encoding function of an edge that emanates from a source node. $n$ is the number of symbols from $\mathcal{F}$ that are transmitted across an edge in the network. Thus for such an edge $e$ whose $\tail (e) = s \in S$, the local encoding function is $\tilde{\phi}_e(X_{s1}, X_{s2}, \ldots, X_{sm}) \in \mathcal{F}^n$. We will be using both row and column vectors in this paper and they will be explicitly mentioned while defining them. %All vectors considered in this paper are assumed to be column vectors unless mentioned otherwise.
If $u$ is a vector, the $u^T$ represents its transpose.

\begin{itemize}
\item Local encoding function for edge $e \in E$.
\begin{align*}
\tilde{\phi}_e &: \mathcal{F}^m \rightarrow \mathcal{F}^n ~~ \text{if tail}(e)\in S,  \\
\tilde{\phi}_e &: \mathcal{F}^{n|\In(\text{tail}(e))|} \rightarrow \mathcal{F}^n ~~ \text{if tail}(e) \notin S.
\end{align*}

\item Decoding function for the terminal $t \in T$.
\begin{align*}
\psi_{t} : \mathcal{F}^{n|\In(t)|} \rightarrow \mathcal{F}^m.
\end{align*}
\end{itemize}

A network code is linear over the finite field $\mathcal{F}$ if all the local encoding and decoding functions are linear transformations over $\mathcal{F}$. In this case the local encoding functions can be represented via matrix products where the matrix elements are from $\mathcal{F}$. For example, for an edge $e$ such that $\tail(e) \notin S$, let $c \in \mathbb{N}$ be such that $c = |\In(\tail(e))|$ and $\In(\tail(e))=\{e_1,e_2,\ldots,e_c\}$. Also, let each $\phi_{e_i}(X) \in \mathcal{F}^n$ be denoted as a column vector of size $n$ whose elements are from $\mathcal{F}$. Then the value transmitted on $e$ can be evaluated as% \aditya{following eq. is not correct}
\begin{IEEEeqnarray*}{Rl}
  \phi_e(X)=&\tilde{\phi}_e(\phi_{e_1}(X),\phi_{e_2}(X), \ldots, \phi_{e_c}(X)),\\
  =&M_e \begin{bmatrix}
         \phi_{e_1}(X)^T & \phi_{e_2}(X)^T & \hdots & \phi_{e_c}(X)^T
       \end{bmatrix}^T,
  %=M_{e,1}\phi_{e_1}(X)+M_{e,2}\phi_{e_2}(X) + \cdots + M_{e,c}\phi_{e_c}(X)
\end{IEEEeqnarray*} where $M_{e} \in \mathcal{F}^{n\times nc}$ is a matrix indicating the local encoding function for edge $e$.
%, i.e., the elements of the matrices that correspond to the transformations are from the finite field $\mathcal{F}$.
For the sum-networks that we consider, a valid $(m,n)$ fractional network code solution over $\mathcal{F}$ has the interpretation that the component-wise sum over $\mathcal{F}$ of $m$ i.i.d. source symbols can be communicated to all the terminals in $n$ time slots.
\begin{definition}
The \textit{rate} of a $(m,n)$ network code is defined to be the ratio $m/n$. A sum-network is solvable if it has a $(m,m)$ network coding solution for some $m \in \mathbb{N}$.
\end{definition}
\begin{definition}
The \textit{computation capacity} of a sum-network is defined as
\begin{equation*}
%\sup \left\lbrace \frac{m}{n}: ~\text{there is a valid}~ (m,n)~\text{network code for the given sum-network.}~\right\rbrace
\sup \left\lbrace \frac{m}{n}: \begin{IEEEeqnarraybox*}[][c]{,t,} there is a valid $(m,n)$ network code\\for the given sum-network.
\end{IEEEeqnarraybox*}\right\rbrace
\end{equation*}
\end{definition}
%The rate of this network code is defined to be $m/n$. A network is said to be solvable if it has a $(m,m)$ network coding solution for some $m \geq 1$. A network is said to have a scalar solution if it has a $(1,1)$ solution. The supremum of all achievable rates is called the capacity of the network.

We use different types of {\it incidence structures} for constructing sum-networks throughout this paper. We now formally define and present some examples of incidence structures.
\begin{definition} {\it Incidence Structure.}
Let $\mathcal{P}$ be a set of elements called \textit{points}, and $\mathcal{B}$ be a set of elements called \textit{blocks}, where each block is a subset of $\mathcal{P}$. The incidence structure $\mathcal{I}$ is defined as the pair $(\mathcal{P}, \mathcal{B})$. %\footnote{In general $\mathcal{B}$ can be a multiset, i.e., it can contain repeated elements, but we will not be considering them in our work}.
If $p \in \mathcal{P}, B \in \mathcal{B}$ such that $p \in B$, then we say that point $p$ is incident to block $B$. In general $\mathcal{B}$ can be a multiset, i.e., it can contain repeated elements, but we will not be considering them in our work. Thus for any two distinct blocks $B_1,B_2$ there is at least one point which is incident to one of $B_1$ and $B_2$ and not the other.
\end{definition}

We denote the cardinalities of the sets $\mathcal{P}$ and $\mathcal{B}$ by the constants $v$ and $b$ respectively. Thus the set of points and blocks can be indexed by a subscript, and we have that
\begin{equation*}
\mathcal{P} = \{p_1,p_2, \ldots,p_v\},~\text{and}~ \mathcal{B}=\{B_1,B_2,\ldots,B_b\}.
\end{equation*}

\begin{definition} {\it Incidence matrix.}
The incidence matrix associated with the incidence structure $\mathcal{I}$ is a $(0,1)$-matrix of dimension $v \times b $ defined as follows.
\begin{equation*}
\setlength{\nulldelimiterspace}{0pt}
A_\mathcal{I}(i,j)\coloneqq\left\lbrace\begin{IEEEeqnarraybox}[\relax][c]{l's}
1 &if $p_i \in B_j$,\\
0 &otherwise.
\end{IEEEeqnarraybox}\right.
\end{equation*}
\end{definition}

Thus, incidence matrices can be viewed as general set systems. For example, a simple undirected graph can be viewed as an incidence structure where the vertices are the points and edges are the blocks (each block is of size two). Combinatorial designs \cite{Stinson} form another large and well-investigated class of incidence structures. In this work we will use the properties of $t$-designs which are defined next.
\begin{definition} {\it $t$-design.}
An incidence structure $\mathcal{I}=(\calP,\calB)$ is a $t$-$(v,k,\lambda)$ design, if
\begin{itemize}
\item it has $v$ points, i.e., $|\calP| =v$,
\item each block $B \in \calB$ is a $k$-subset of the point set $\calP$, and
\item $\calP$ and $\calB$ satisfy the $t$-\textit{design property}, i.e., any $t$-subset of $\calP$ is present in exactly $\lambda$ blocks.
\end{itemize}
\end{definition}
A $t$-$(v,k,\lambda)$ design is called \textit{simple} if there are no repeated blocks. %, i.e., $\calB$ is a set and not a multiset. In this work, we work exclusively with simple $t$-$(v,k,\lambda)$ designs.
These designs have been the subject of much investigation when $t=2$; in this case they are also called balanced incomplete block designs (BIBDs).
\begin{example}
A famous example of a $2$-design with $\lambda = 1$ is the Fano plane $\mathcal{I}=(\calP,\calB)$ shown in Figure \ref{fig:fano}.
\begin{figure}
  \centering
  \includegraphics[width=0.3\textwidth]{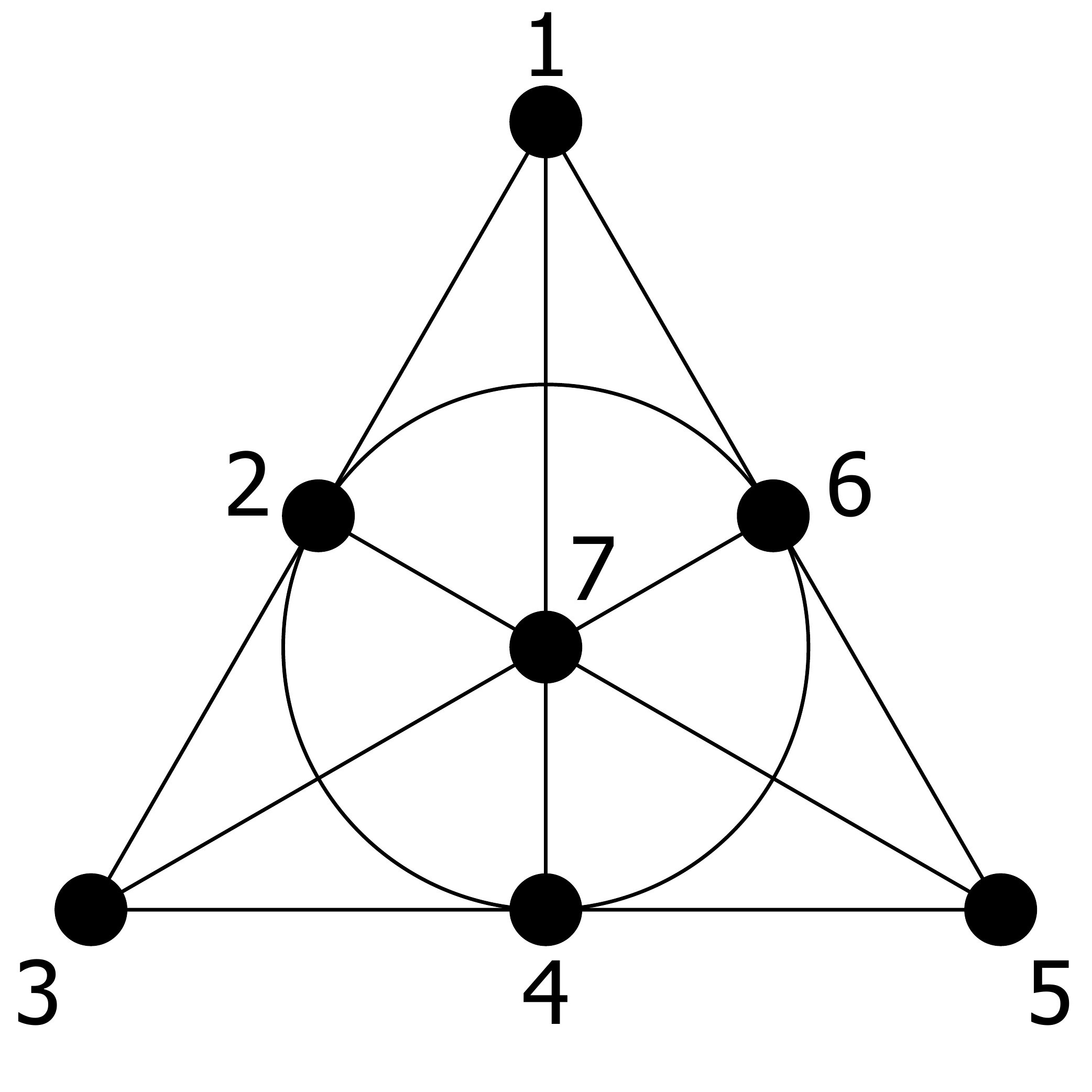}
  \caption{\label{fig:fano} A pictorial depiction of the Fano plane. The point set $\calP = \{1, \dots, 7\}$. The blocks are indicated by a straight line joining their constituent points. The points $2,4$ and $6$ lying on the circle also depict a block.}
\end{figure}
 Letting numerals denote points and alphabets denote blocks for this design, we can write:
\begin{IEEEeqnarray*}{C}
\calP=\{1,2,3,4,5,6,7\},\calB=\{A,B,C,D,E,F,G\}, ~\text{where}\\
A=\{1,2,3\}, B=\{3,4,5\},
C=\{1,5,6\},D=\{1,4,7\},\\E=\{2,5,7\}, F=\{3,6,7\},G=\{2,4,6\}.
\end{IEEEeqnarray*}
%It can be verified that every point $p \in P$ is present in $r=\frac{\lambda (v-1)}{k-1}=\frac{1(7-1)}{3-1}=3$ blocks.
%We let the natural ordering of numerals and alphabets specify the ordering of points and blocks in this design.
The corresponding incidence matrix $A_{\mathcal{I}}$, with rows and columns arranged in numerical and alphabetical order, is shown below.
\begin{equation}\label{eq:Afano}
A_{\mathcal{I}}=
\begin{bmatrix}
1 & 0 & 1 & 1 & 0 & 0 & 0\\1 & 0 & 0 & 0 & 1 & 0 & 1\\1 & 1 & 0 & 0 & 0 & 1 & 0\\0 & 1 & 0 & 1 & 0 & 0 & 1\\0 & 1 & 1 & 0 & 1 & 0 & 0\\0 & 0 & 1 & 0 & 0 & 1 & 1\\0 & 0 & 0 & 1 & 1 & 1 & 0
\end{bmatrix}.
\end{equation}
It can be verified that every pair of points in $\calP$ appears in exactly one block in $\mathcal{B}$.
\vspace{2mm}
\end{example}

There are some well-known conditions that the parameters of a $t$-$(v,k,\lambda)$ design satisfy (see \cite{Stinson}). %In particular if $\calP$ and $\calB$ denote the set of points and blocks, then we have the following
\begin{itemize}
\item For integer $i\leq t$ the number of blocks incident to any $i$-subset of $\calP$ is the same. We let $b_i$ denote that constant. Then,%number of blocks that are incident to any $i$-subset of $\calP$ where $i \leq t$. Then,
\begin{equation}\label{eq:t_design_b_i}
b_i = \lambda \binom{v-i}{t-i}/\binom{k-i}{t-i}, \:\: \forall i \in \{0,1,2,\ldots,t\}.
\end{equation}
We note that $b_0$ is simply the total number of blocks denoted by $b$. Likewise, $b_1$ represents the number of blocks that each point is incident to; we use the symbol $\rho$ to represent it. Furthermore, $b_t = \lambda$.

It follows that a necessary condition for the existence of a $t$-$(v,k,\lambda)$ design is that $\binom{k-i}{t-i}$ divides $\lambda\binom{v-i}{t-i}$ for all $i = 1,2,\ldots, t$.
\item Counting the number of ones in the point-block incidence matrix for a particular design in two different ways, we arrive at the equation $bk=v\rho$.
\end{itemize}

%\section{Summary of Results}
%TBD
%Need to discuss main results and the significance of the results.

\section{Construction of a family of sum-networks}
\label{sec:constr}
Let $[t]\coloneqq\{1,2,\ldots,t\}$ for any $t \in \mathbb{N}$. %For two column vectors $u$ and $v$, denote their inner product by $u^Tv$.
Our construction takes as input a $(0,1)$-matrix $A$ of dimension $r \times c$.
\begin{definition}{\it Notation for row and column of $A$.}
  Let $\bm{p}_i$ denote the $i$-th row vector of $A$ for $i \in [r]$ and $\bm{B}_j$ denote the $j$-th column vector of $A$ for $j \in [c]$ \footnote{A justification for this notation is that later when we use the incidence matrix ($A_{\mathcal{I}}$) of an incidence structure $\mathcal{I}$ to construct a sum-network, the rows and columns of the incidence matrix will correspond to the points and blocks of $\mathcal{I}$ respectively.}.
\end{definition}
%We explain our construction of sum-networks by referring to a $(0,1)$-matrix $A$ of dimension $r \times c$.
It turns out that the constructed sum-networks have interesting properties when the matrix $A$ is the incidence matrix of appropriately chosen incidence structures. The construction algorithm is presented in Algorithm \ref{alg:constr}. The various steps in the algorithm that construct components of the sum-network $G=(V,E)$ are described below.
\begin{enumerate}
  \item \textit{Source node set $S$ and terminal node set $T$:} $S$ and $T$ both contain $r+c$ nodes, one for each row and column of $A$. The source nodes are denoted at line \ref{alg:constr_S} as $s_{p_i},s_{B_j}$ if they correspond to the $i$-th row, $j$-th column respectively. The terminal nodes are also denoted in a similar manner at line \ref{alg:constr_T}. They are added to the vertex set $V$ of the sum-network at line \ref{alg:constr_V}.
  \item \textit{Bottleneck edges:} We add $r$ unit-capacity edges indexed as $e_i$ for $i\in [r]$ in line \ref{alg:constr_bottleneck} to the edge set $E$. Each edge $e_i$ corresponds to a row of the matrix $A$. We also add the required tail and head vertices of these edges to $V$.
  \item \textit{Edges between $S \cup T$ and the bottleneck edges:} For every $i \in [r]$, we connect $\tail(e_i)$ to the source node corresponding to the row $\bm{p}_i$ and to the source nodes that correspond to all columns of $A$ with a $1$ in the $i$-th row. This is described in line \ref{alg:constr_tail} of the algorithm. Line \ref{alg:constr_head} describes a similar operation used to connect each $\head(e_i)$ to certain terminal nodes.
  \item \textit{Direct edges between $S$ and $T$:} For each terminal in $T$, these edges directly connect it to source nodes that do not have a path to that particular terminal through the bottleneck edges. Using the notation for rows and columns of the matrix $A$, they can be characterized as in lines \ref{alg:constr_tp} and \ref{alg:constr_tB}.
\end{enumerate}

\begin{algorithm}
\caption{SUM-NET-CONS}\label{alg:constr}
\begin{algorithmic}[1]
\REQUIRE $A$.% \COMMENT{$A$ is a $(0,1)$-matrix of size $r \times c$}
\ENSURE $G=(V,E)$.% \COMMENT{$G$ is the directed acyclic sum-network returned}
\STATE Initialize $V,E,S,T \leftarrow \phi$.
\STATE $E \leftarrow \{e_i : i \in [r]\}$.\label{alg:constr_bottleneck}% \COMMENT{these are bottleneck edges}
\STATE $V \leftarrow \{\head(e_i),\tail(e_i):i \in [r]\}$.
\STATE $S \leftarrow \{s_{p_i}:i \in [r]\}\cup\{s_{B_j}:j \in [c]\}$.\label{alg:constr_S}% \COMMENT{the source-node set}
\STATE $T \leftarrow \{t_{p_i}:i \in [r]\} \cup \{t_{B_j}:j \in [c]\}$.\label{alg:constr_T}% \COMMENT{the terminal-node set}
\STATE $V \leftarrow V \cup S \cup T$.\label{alg:constr_V}
\FORALL{$i \in [r]$}%[edges connecting the bottleneck edges to source and terminal nodes]
\STATE $E \leftarrow E \cup \{(s_{B_j},\tail(e_i)):A(i,j)=1; j \in [c]\} \cup \{(s_{p_i},\tail(e_i))\}$.\label{alg:constr_tail}
\STATE $E \leftarrow E \cup \{(\head(e_i), t_{B_j}):A(i,j)=1; j \in [c]\} \cup \{(\head(e_i),t_{p_i})\}$.\label{alg:constr_head}
\ENDFOR
\FORALL{$i \in [r]$}%[direct edges connecting source nodes to row-terminals]
\STATE $E \leftarrow E \cup \{(s_{p_j},t_{p_i}): i \neq j; j \in[r]\} \cup \{(s_{B_j},t_{p_i}):A(i,j)=0; j \in [c]\}$. \label{alg:constr_tp}
\ENDFOR
\FORALL{$j \in [c]$}%[direct edges connecting source nodes to column-terminals]
\STATE $E \leftarrow E \cup \{(s_{p_i},t_{B_j}):A(i,j)=0; i \in [r]\} \cup \{(s_{B_{j'}},t_{B_j}): \bm{B}_{j}^T\bm{B}_{j'}=0; j' \in [c]\}$. \label{alg:constr_tB}
\ENDFOR
\RETURN $G \leftarrow (V,E)$.
\end{algorithmic}
\end{algorithm}

For an incidence structure $\mathcal{I}$, let $A_{\mathcal{I}}$ represent its incidence matrix. The sum-networks constructed in the paper are such that the matrix $A$ used in the \sumnetalg algorithm is either equal to $A_{\mathcal{I}}$ or $A_{\mathcal{I}}^T$ for some incidence structure $\mathcal{I}$. When $A = A_\mathcal{I}$, we call the sum-network constructed as the \textit{normal} sum-network for $\mathcal{I}$. Otherwise when $A = A_\mathcal{I}^T$, we call the sum-network constructed as the \textit{transpose} sum-network for $\mathcal{I}$.
The following definitions are useful for analysis.
For every $p \in \mathcal{P}$, we denote the set of blocks that contain the point $p$ as
\begin{equation}\label{eq:<p>}
\langle p\rangle \coloneqq \{B \in \mathcal{B}: p \in B\},
\end{equation} and for every $B \in \mathcal{B}$, the collection of blocks that have a non-empty intersection with $B$ is denoted by the set
\begin{align}\label{eq:<B>}
\langle B\rangle &\coloneqq \{B' \in \mathcal{B} : B' \cap B \neq \phi\} \\
&= \{B' \in \calB:\bm{B}^T\bm{B'}\neq 0\},
\end{align} where boldface $\bm{B}$ indicates the column of $A_\mathcal{I}$ corresponding to block $B \in \mathcal{B}$.

The inner product above is computed over the reals. In the sequel, we will occasionally need to perform operations similar to the inner product over a finite field. This shall be explicitly pointed out.

\begin{figure}[t]
\begin{center}
\subfigure[]{\label{fig:K2normal}
\includegraphics[scale=0.8]{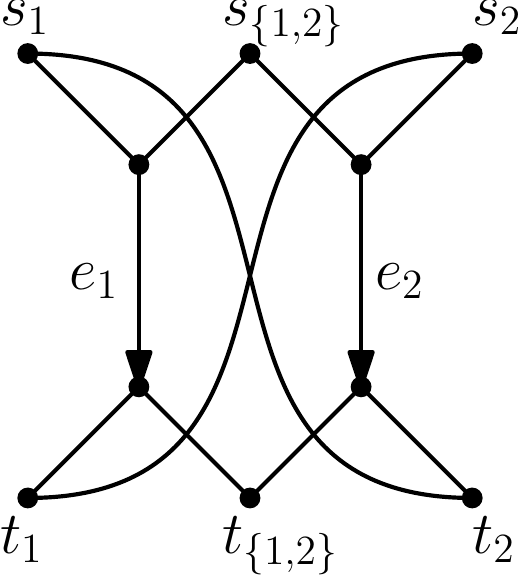}}
\hspace{0.2in}
\subfigure[]{\label{fig:K2transpose}
\includegraphics[scale=0.8]{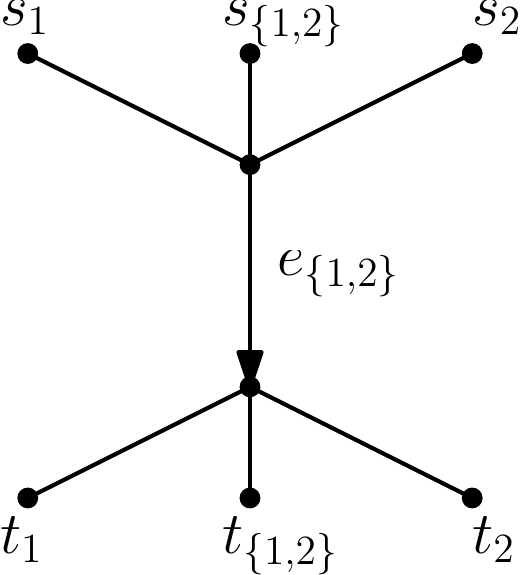}}
\caption{\label{fig:K2} Two sum-networks obtained from the line graph on two vertices described in Example \ref{eg:K2}. The source set $S$ and the terminal set $T$ contain three nodes each. All edges are unit-capacity and point downward. The edges with the arrowheads are the bottleneck edges constructed in step 2 of the construction procedure. (a) Normal sum-network, and (b) transposed sum-network.}
\end{center}
\end{figure}

We now present some examples of sum-networks constructed using the above technique.
\begin{example}\label{eg:K2}
Let $\mathcal{I}$ be the unique simple line graph on two vertices, with points corresponding to the vertices and blocks corresponding to the edges of the graph. Denoting the points as natural numbers, we get that $\mathcal{P}=\{1,2\}$ and $\mathcal{B}=\{\{1,2\}\}$. Then the associated incidence matrices are as follows.
\begin{equation*}
A_\mathcal{I} = \begin{bmatrix}
1\\1
\end{bmatrix},\text{and~} A_\mathcal{I}^T = \begin{bmatrix}
1 & 1
\end{bmatrix}.
\end{equation*} Following the \sumnetalg algorithm the two sum-networks obtained are as shown in the Figure \ref{fig:K2}.
\vspace{2mm}
\end{example}

\begin{example} \label{eg:bibd}
In this example we construct a sum-network using a simple $t$-design. Let $\mathcal{I}$ denote the $2$-$(3,2,1)$ design with its points denoted by the numbers $\{1,2,3\}$ and its blocks denoted by the letters $\{A,B,C\}$. For this design we have that
$A=\{1,2\}, B= \{1,3\}, C=\{2,3\}$
and its associated incidence matrix under row and column permutations can be written as follows.
\begin{equation*}
A_{\mathcal{I}}=\begin{bmatrix}
1 & 1 & 0\\
1 & 0 & 1\\
0 & 1 & 1
\end{bmatrix}
\end{equation*}
Note that $A_{\mathcal{I}}=A_{\mathcal{I}}^T$. Hence the normal sum-network and the transposed sum-network are identical in this case. Following the \sumnetalg algorithm, we obtain the sum-network shown in Figure \ref{fig:bibd}.
\begin{figure}
\centering
\includegraphics[scale=1]{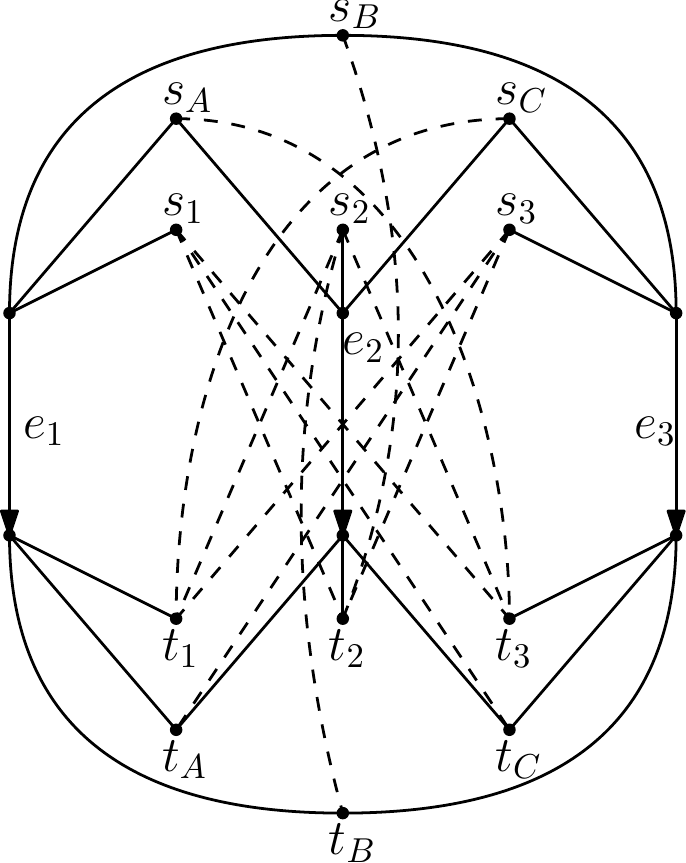}
\caption{The normal sum-network obtained for the incidence structure $\mathcal{I}$ described in Example \ref{eg:bibd}. All edges are unit-capacity and directed downward. The edges with the arrowheads are the bottleneck edges, and the edges denoted by dashed lines correspond to the direct edges introduced in step 4 of the construction procedure. For this case, the normal and the transposed sum-network are identical.}\label{fig:bibd}
\end{figure}
\vspace{2mm}
\end{example}
\begin{remark}
Note that each edge added in the \sumnetalg algorithm has unit capacity. Proposition \ref{prop:multi_cap_edge} in Section \ref{sec:discussion} modifies the \sumnetalg algorithm so that each edge $e$ in the sum-network has $\capacity(e)=\alpha>1, \alpha \in \mathbb{N}$.% where $\alpha \in \mathbb{N}, \alpha > 1$.
\end{remark}
%, i.e., $Z\coloneqq\sum_{i \in [r]} X_{p_i}+\sum_{ j \in [c]} X_{B_j}$.
\section{Upper bound on the computation capacity}\label{sec:upper_bd}
In this section, we describe an upper bound on the computation capacity of a sum-network obtained from a $(0,1)$-matrix $A$ of dimension $r \times c$.
We assume that there exists a $(m,n)$ fractional network code assignment, i.e., $\tilde{\phi}_e$ for $e \in E$ (and corresponding global encoding functions $\phi_e(X)$) and decoding functions $\psi_t$ for $t \in T$ so that all the terminals in $T$ can recover the sum of all the independent sources.

For convenience of presentation, we will change notation slightly and let the messages observed at the source nodes corresponding to the rows of $A$ as $X_{p_i}$ for $i \in [r]$ and those corresponding to the columns of $A$ as $X_{B_j}$ for $j \in [c]$. Each of the messages is a column vector of length $m$ over $\mathcal{F}$. The set of all source messages is represented by $X$. %For the $(m,n)$ fractional network code, $X_{p_i}, X_{B_j} \in \mathcal{F}^m$ for all $i \in [r], j \in [c]$ respectively and $Z \in \mathcal{F}^m$ denotes the component-wise sum of all the source messages.
%In what follows, we will use the following notation.
We let $\phi_e(X)$ denote the $n$-length column vector of symbols from $\mathcal{F}$ that are transmitted by the edge $e \in E$, as it is the value returned by the global encoding function $\phi_e$ for edge $e$ on the set of source messages denoted by $X$. As is apparent, non-trivial encoding functions can only be employed on the bottleneck edges, i.e., $e_i$ for $i \in [r]$ as these are the only edges that have more than one input. For brevity, we denote $\phi_i(X) = \phi_{e_i}(X)$. We define the following set of global encoding functions.
\begin{IEEEeqnarray*}{Rl}
\phi_{\In(v)}(X)\coloneqq &\{\phi_e(X): e \in \In (v)\}, ~\forall v \in V.% \text{~and}\\
%\phi_{\In (e_i)}(X)\coloneqq &\{\phi_e(X): \head(e) = \tail (e_i); e \in E\}, ~\forall i \in [r].
\end{IEEEeqnarray*}
Let $H(Y)$ be the entropy function for a random variable $Y$. We let $\{Y_i\}_{1}^{l}$ denote the set $\{Y_1,Y_2,\ldots, Y_l\}$ for any $l > 1$.
The following lemma demonstrates that certain partial sums can be computed by observing subsets of the bottleneck edges.
\begin{lemma}\label{lem:partial_sums}
If a network code allows each terminal to compute the demanded sum, then the value $X_{p_i}' \coloneqq X_{p_i} + \sum_{j:A(i,j)=1}X_{B_j}$ can be computed from $\phi_i(X)$, i.e., $H\left(X_{p_i}' | \phi_{i}(X)\right)=0$ for all $i \in [r]$. Similarly for any $j \in [c]$ the value $X_{B_j}' \coloneqq \sum_{i:A(i,j)=1}X_{p_i}+\sum_{j':B_{j'} \in \langle B_j\rangle}X_{B_{j'}}$ can be computed from the set of values $\{\phi_i(X): \text{~for~} i\in [r], A(i,j)=1\}$.
\end{lemma}
\begin{IEEEproof}
We let for any $i \in [r]$
\begin{equation*}
Z_1 = \sum_{i' \neq i} X_{p_{i'}},~
Z_2 = \sum_{j: A(i,j)=1} X_{B_j} ~\text{and}~
Z_3 = \sum_{j: A(i,j)=0} X_{B_j},
\end{equation*}
such that the sum $Z= X_{p_i}+ Z_1 + Z_2 + Z_3$ and $X'_{p_i}=X_{p_i}+Z_2$.

By our assumption that each terminal can recover the demanded sum, we know that $Z$ can be evaluated from $\phi_{\In\left(t_{p_i}\right)}(X)$ for all $i \in [r]$, i.e.,
$ H\left( Z | \phi_{\In(t_{p_i})}(X)\right)=0$ for all $i \in [r]$.
Since $\{X_{p_{i'}} : i' \neq i\}$ and $\{X_{B_j} : A(i,j)=0\}$ determine the value of $Z_1$ and $Z_3$ respectively and also determine the values transmitted on each of the direct edges that connect a source node to $t_{p_i}$, we get that
%{\allowdisplaybreaks
\begin{IEEEeqnarray*}{Rl}
H&\left( Z | \phi_{\In\left(t_{p_i}\right)}(X)\right)\\
=&H\left( Z | \phi_{i}(X), \{\phi_{(s_{p_{i'}},t_{p_i})}(X):i'\neq i\},\right.\\
& \hfill\left.\{\phi_{(s_{B_j},t_{p_i})}: A(i,j)=0\}\right)\\
\stackrel{\mathrm{(a)}}{\geq} & H\left( X_{p_i}+Z_1+Z_2+Z_3 | \phi_{i}(X), \{X_{p_{i'}} : i' \neq i\},\right.\\
& \hfill\left.\{X_{B_j} : A(i,j)=0\}\right)\\
=&H\left( X_{p_i}' | \phi_{i}(X), \{X_{p_{i'}} : i' \neq i\},\{X_{B_j} : A(i,j)=0\}\right)\\
=&H\left(X_{p_i}',\{X_{p_{i'}} : i' \neq i\},\{X_{B_j} : A(i,j)=0\}|\phi_{i}(X)\right)\\
&-\: H\left(\{X_{p_{i'}} : i' \neq i\},\{X_{B_j} : A(i,j)=0\}|\phi_{i}(X)\right)\\
=&H\left(X_{p_i}'|\phi_{i}(X)\right)\\
&+\:H\left(\{X_{p_{i'}} : i' \neq i\},\{X_{B_j} : A(i,j)=0\}|X_{p_i}',\phi_{i}(X)\right)\\
&-\: H\left(\{X_{p_{i'}} : i' \neq i\},\{X_{B_j} : A(i,j)=0\}|\phi_{i}(X)\right)\\
\stackrel{\mathrm{(b)}}{=} &H\left(X_{p_i}' | \phi_{i}(X)\right), \IEEEyesnumber \label{eq:decode_t_p}
\end{IEEEeqnarray*}%}
where inequality $(\mathrm{a})$ follows from the fact that $\phi_{(s_{p_{i'}},t_{p_i})}(X)$ is a function of $X_{p_{i'}}$ for $i' \neq i$ and $\phi_{(s_{B_j},t_{p_i})}(X)$ is a function of $\{X_{B_j}: A(i,j) = 0\}$ %(similarly for $\phi_{(s_{B_j},t_{p_i})}(X)$) % conditioning reduces entropy
and equality $(\mathrm{b})$ is due to the fact that $X'_{p_i}$ is conditionally independent of both $\{X_{p_{i'}}:i'\neq i\}$ and $\{X_{B_j}:A(i,j)=0\}$ given $\phi_i(X)$. This conditional independence can be checked as follows. Let bold lowercase symbols represent specific realizations of the random variables.
%{\allowdisplaybreaks
\begin{IEEEeqnarray*}{rL}
&\Pr\left(X_{p_i}'=\bm{x}_{p_i}',\{X_{p_{i'}}=\bm{x}_{p_{i'}} : i' \neq i\},\right.\\
&\hfill \left.\{X_{B_j}=\bm{x}_{B_j} : A(i,j)=0\}|\phi_{i}(X) = \phi_i (\bm{x})\right)\\
\stackrel{\mathrm{(a)}}{=}&\Pr(X_{p_i}'=\bm{x}_{p_i}',\phi_{i}(X)=\phi_i(\bm{x}))/\Pr(\phi_{i}(X)=\phi_i(\bm{x}))\\
&\:\cdot \Pr(\{X_{p_{i'}}=\bm{x}_{p_{i'}} : i' \neq i\},\{X_{B_j}=\bm{x}_{B_j} : A(i,j)=0\})\\
%&=\Pr\left(X_{p_i}'=\bm{x}_{p_i}'|\phi_{i}(X)=\phi_i(\bm{x})\right)\Pr(\{X_{p_{i'}}=\bm{x}_{p_{i'}} : i' \neq i\},\{X_{B_j}=\bm{x}_{B_j} : A(i,j)=0\})\\
\stackrel{\mathrm{(b)}}{=}&\Pr\left(X_{p_i}'=\bm{x}_{p_i}'|\phi_{i}(X)=\phi_i(\bm{x})\right)\\
&\:\cdot \Pr\left(\{X_{p_{i'}}=\bm{x}_{p_{i'}} : i' \neq i\},\{X_{B_j}=\bm{x}_{B_j} : A(i,j)=0\}\right.\\
&\hfill \left.\vphantom{\{X_{B_j}=\bm{x}_{B_j} : A(i,j)=0\}}|\phi_{i}(X)=\phi_i(\bm{x})\right),
\end{IEEEeqnarray*}%}
where equalities $(\mathrm{a})$ and $(\mathrm{b})$ are due to the fact that the source messages are independent and $\phi_i(\bm{x})$ is only a function of $\bm{x}_{p_i}$ and the set $\{\bm{x}_{B_j}:A(i,j)=1\}$.

%\begin{equation*}
%\Pr\left(X_{r_i}',\{X_{r_{i'}} : i' \neq i\},\{X_{c_j} : A(i,j)=0\}|\phi_{i}(X)\right)=\frac{\Pr(X_{r_i}',\phi_{i}(X))\Pr(\{X_{r_{i'}} : i' \neq i\},\{X_{c_j} : A(i,j)=0\})}{\phi_{i}(X)} = \Pr\left(X_{r_i}'|\phi_{i}(X)\right)\Pr(\{X_{r_{i'}} : i' \neq i\},\{X_{c_j} : A(i,j)=0\}) = \Pr\left(X_{r_i}'|\phi_{i}(X)\right)\Pr(\{X_{r_{i'}} : i' \neq i\},\{X_{c_j} : A(i,j)=0\}|\phi_{i}(X)).
%\end{equation*}
Since terminal $t_{p_i}$ can compute $Z$, $H\left( Z | \phi_{\In\left(t_{p_i}\right)}(X)\right)=0$ and we get from eq. \eqref{eq:decode_t_p} that $H(X_{p_i}+Z_2 | \phi_{i}(X))=0$.

For the second part of the lemma, we argue similarly as follows. % \aditya{include changes similar to above}
We let for any $j \in [c]$
\begin{IEEEeqnarray*}{C}
Z_1=\sum_{i:A(i,j)=1} X_{p_i}, Z_2=\sum_{i:A(i,j)=0} X_{p_i},\\
Z_3=\sum_{B\in \langle B_{j} \rangle}X_{B}, Z_4=\sum_{B \notin \langle B_{j}\rangle}X_{B}
\end{IEEEeqnarray*}
such that $Z= Z_1+Z_2+Z_3+Z_4$ and $X_{B_j}'=Z_1+Z_3$.
By our assumption, for all $j \in [c]$, $H\left(Z | \phi_{\In\left(t_{B_j}\right)}(X)\right)=0$.
The sets $ \{X_p : p \notin B_j\}$ and $ \{X_B : B \notin \langle B_j \rangle\}$ determine the value of $Z_2$ and $Z_4$ respectively and also the values transmitted on each of the direct edges that connect a source node to the terminal $t_{B_j}$. Let $\Phi$ denote the set $\{\phi_i(X):A(i,j)=1\}$. Then,%Hence, we have that
\begin{IEEEeqnarray*}{Rl}
H&\left( Z | \phi_{\In\left(t_{B_j}\right)}(X)\right)\\
=&H\left(Z_1+Z_2+Z_3+Z_4 | \Phi, \{\phi_{(s_{p_i},t_{B_j})}(X)\!:\! A(i,j)=0\},\right.\\
 & \hfill \left.\{\phi_{(s_{B},t_{B_j})}:B \notin \langle B_{j}\rangle \}\right)\\
\stackrel{\mathrm{(a)}}{\geq}& H\left(Z_1+Z_2+Z_3+Z_4 | \Phi, \{X_{p_i}:A(i,j)=0\},\right.\\
& \hfill \left. \{X_{B}:B \notin \langle B_{j}\rangle \}\right)\\
=& H\left( X_{B_j}' | \Phi, \{X_{p_i}:A(i,j)=0\},\{X_{B}:B\notin \langle B_{j}\rangle\}\right)\\
=& H\left(X_{B_j}',\{X_{p_{i}}:A(i,j)=0\},\{X_{B}:B\notin \langle B_{j}\rangle\}|\Phi\right)\\
 & \hfill -\: H\left(\{X_{p_{i}}:A(i,j)=0\},\{X_{B}:B\notin \langle B_{j}\rangle\}|\Phi\right)\\
=& H(X_{B_j}'|\Phi)-\! H (\{X_{p_i}\! :\! A(i,j)=0\},\{X_{B}\!: \! B\notin \langle B_{j}\rangle\}|\Phi)\\
 & \hfill +\: H(\{X_{p_i}:A(i,j)=0\},\{X_{B}:B\notin \langle B_{j}\rangle\}|X_{B_j}',\Phi)\\
\stackrel{\mathrm{(b)}}{=}&H(X_{B_j}' |\Phi).
\end{IEEEeqnarray*}%}
Inequality $(\mathrm{a})$ is due to the fact that $\phi_{(s_{p_i},t_{B_j})}(X)$ is a function of $X_{p_i}$ and similarly for $\phi_{(s_{B},t_{B_j})}(X)$. Equality $(\mathrm{b})$ follows from the fact that $Z_1+Z_3$ is conditionally independent of both $\{X_{p_i}:A(i,j)=0\}$ and $\{X_{B_{j'}}:B\notin \langle B_{j}\rangle\}$ given the set of random variables $\{\phi_{i}(X) : A(i,j)=1\}$. This can be verified in a manner similar to as was done previously. This gives us the result that $H(X_{B_j}' | \{\phi_{i}(X) : A(i,j)=1\})=0$.
\end{IEEEproof}

Next, we show the fact that the messages observed at the source nodes are independent and uniformly distributed over $\mathcal{F}^m$ imply that the random variables $X_{p_i}'$ for all $i \in [r]$ are also uniform i.i.d. over $\mathcal{F}^m$. To do that, we introduce some notation. For a matrix $N \in \mathcal{F}^{r \times c}$, for any two index sets $\mathcal{R} \subseteq [r], \mathcal{C} \subseteq [c]$, we define the submatrix of $N$ containing the rows indexed by $\mathcal{R}$ and the columns indexed by $\mathcal{C}$ as $N[\mathcal{R},\mathcal{C}]$. Consider two $(0,1)$-matrices $N_1,N_2$ of dimensions $r_1 \times t$ and $t \times c_2$ respectively. Here $1$ and $0$ indicate the multiplicative and additive identities of the finite field $\mathcal{F}$ respectively. The $i$-th row of $N_1$ is denoted by the row submatrix $N_1\left[i,[t]\right] \in \{0,1\}^t$ and the $j$-th column of $N_2$ be denoted by the column submatrix $N_2\left[[t],j\right] \in \{0,1\}^t$. Then we define a matrix function on $N_1N_2$ that returns a $r_1 \times c_2$ matrix $(N_1N_2)_\#$ as follows.
\begin{equation*}
  (N_1N_2)_\#(i,j) = \begin{cases}
                   1, & \begin{IEEEeqnarraybox*}[][c]{s,} if the product $N_1\left[i,[t]\right]N_2\left[[t],j\right]$\\over $\mathbb{Z}$ is positive,
                        \end{IEEEeqnarraybox*}\\
                   %\mbox{if the inner product } N_1\left[i,[t]\right]N_2\left[[t],j\right] \mbox{ over } \mathbb{Z} \mbox{ is positive} \\
                   0, & \mbox{otherwise}.
                 \end{cases}
\end{equation*}
For an incidence structure $\mathcal{I}=(\calP,\calB)$ with $r \times c$ incidence matrix $A$, let $X_p, ~\forall p \in \calP$ and $X_B, ~ \forall B \in \calB$ be $m$-length vectors with each component i.i.d. uniformly distributed over $\mathcal{F}$. We collect all the independent source random variables in a column vector $\bm{X}$ having $m(r+c)$ elements from $\mathcal{F}$ as follows
\begin{equation*}
\bm{X}\mathrel{\mathop:}=\begin{bmatrix}
X_{p_1}^T & X_{p_2}^T & \cdots & X_{p_r}^T & X_{B_1}^T & X_{B_2}^T & \cdots & X_{B_c}^T
\end{bmatrix}^T.
\end{equation*}
Recall that $\bm{p}_i$ denotes the $i$-th row and $\bm{B}_j$ denotes the $j$-th column of the matrix $A$. For all $i \in [r]$ let $\bm{e}_i \in \mathcal{F}^r$ denote the column vector with $1$ in its $i$-th component and zero elsewhere.
Then for $X_{p_i}', X_{B_j}'$ as defined in lemma \ref{lem:partial_sums}, one can check that ($\otimes$ indicates the Kronecker product of two matrices)
\begin{IEEEeqnarray}{Rl}
X_{p_i}^{'}=&\left(\begin{bmatrix}
\bm{e}_i^T & \bm{p}_i
\end{bmatrix}\otimes I_m\right)\bm{X}, \;\; \text{for all}\; i \in [r]\;\; \text{and} \label{eq:X_p_i'}\\
X_{B_j}^{'}=&\left(\begin{bmatrix}
\bm{B}_j^T & (\bm{B}_j^T\bm{B}_1)_\# & \ldots & (\bm{B}_j^T\bm{B}_c)_\#
\end{bmatrix}\otimes I_m\right) \bm{X},\IEEEeqnarraynumspace \label{eq:X_B_j'}%& (\bm{B}_j^T\bm{B}_2)_\# %, \;\; \text{for all}\; j \in [c]
\end{IEEEeqnarray}for all $j \in [c]$ where $I_m$ is the identity matrix of size $m$.
By stacking these values in the correct order, we can get the following matrix equation.
\begin{equation}\label{eq:M_A_eqn}
  \begin{bmatrix}
    X_{p_1}^{'T} & \cdots & X_{p_r}^{'T} & X_{B_1}^{'T} & \cdots & X_{B_c}^{'T}
  \end{bmatrix}^T = (M_A \otimes I_m) \bm{X}% & X_{p_2}^{'T} %  & X_{B_2}^{'T}
\end{equation} where the matrix $M_A \in \mathcal{F}^{(r+c) \times (r+c)}$ is defined as
\begin{equation}\label{eq:defn_M_A}
  M_A \coloneqq \begin{bmatrix}
           I_r & A \\
           A^T & (A^TA)_\#
         \end{bmatrix}.
\end{equation} Note that the first $r$ rows of $M_A$ are linearly independent. There is a natural correspondence between the rows of $M_A$ and the points and blocks of $\mathcal{I}$ of which $A$ is the incidence matrix. If $1 \leq i \leq r$, then the $i$-th row $M_A\left[i,[r+c]\right]$ corresponds to the point $p_i \in \calP$ and if $r+1 \leq j \leq r+c$, then the $j$-th row $M_A\left[j,[r+c]\right]$ corresponds to the block $B_j \in \calB$.
\begin{lemma}\label{lem:partial_sums_indep}
For a $(0,1)$-matrix $A$ of size $r \times c$, let $X_{p_i}', X_{B_j}' \in \mathcal{F}^m$ be as defined in Eqs. \eqref{eq:X_p_i'}, \eqref{eq:X_B_j'} and matrix $M_A$ be as defined in eq. \eqref{eq:defn_M_A}. Let $r + t \coloneqq \rank_{\mathcal{F}}M_A$ for some non-negative integer $t$ and index set $\mathcal{S}' \subseteq \{r+1,r+2,\ldots,r+c\}$ be such that $\rank_{\mathcal{F}} M_A\left[[r]\cup\mathcal{S}',[r+c]\right]=r+t$. Let $\calB_{\mathcal{S}'}\coloneqq\{B_{\mathcal{S}'_1},B_{\mathcal{S}'_2},\ldots,B_{\mathcal{S}'_t}\} \subseteq \calB$ be the set of blocks that correspond to the rows of $M_A$ indexed by $\mathcal{S}'$ in increasing order. Then we have
\begin{IEEEeqnarray}{C}
  \Pr\left(X_{p_1}'=x_1',\ldots,X_{p_r}'=x_r',X_{B_{\mathcal{S}'_1}}'=y_1',\ldots, X_{B_{\mathcal{S}'_t}}'=y_t'\right)\IEEEnonumber\\
  \hfill = q^{-m(r+t)},%\prod_{i=1}^{r}\Pr\left(X_{p_i}'=x_i'\right)\prod_{j=1}^{t}\Pr\left(X_{B_{\mathcal{S}'_j}}=y_j'\right),
  ~\text{and}
  \IEEEeqnarraynumspace \label{eq:psums_indep}\\
  \Pr\left(X_{p_i}'=x_i'\right)=\Pr\left(X_{B_{\mathcal{S}'_j}}'=y_j'\right)=q^{-m}, \forall i \in [r], j \in [t].
  \IEEEnonumber
\end{IEEEeqnarray}
\end{lemma}
\begin{IEEEproof}
The quantities in the statement of the lemma satisfy the following system of equations
\begin{IEEEeqnarray*}{C}%\label{eq:app_proof}
\left(M\! \left[[r]\cup \mathcal{S}'\! ,[r+c]\right]\otimes I_m\right)\!
\begin{bmatrix}
X_{p_1}^T & \!\!\!\cdots \!\!\! & X_{p_r}^T \!\!\!& X_{B_1}^T & \!\!\!\cdots \!\!\! & X_{B_c}^T
\end{bmatrix}^T\\
=
\begin{bmatrix}
X_{p_1}^{'T} & \!\!\!\cdots \!\!\!& X_{p_r}^{'T} & X_{B_{\mathcal{S}'_1}}^{'T} & \!\!\!\cdots \!\!\!& X_{B_{\mathcal{S}'_t}}^{'T}
\end{bmatrix}^T.
\end{IEEEeqnarray*}
The vector $\begin{bmatrix} X_{p_1}^T & \cdots & X_{p_r}^T & X_{B_1}^T & \cdots & X_{B_c}^T \end{bmatrix}^T$ is uniform over $\mathcal{F}^{m(r+c)}$.
Since the matrix $M\left[[r]\cup \mathcal{S}',[r+c]\right]\otimes I_m$ has full row rank equal to $m(r+t)$, the R.H.S. of the above equation is uniformly distributed over $\mathcal{F}^{m(r+t)}$, giving the first statement. The second statement is true by marginalization.
\end{IEEEproof}

%The above lemmas are useful in stating the first major result of this paper.
\begin{theorem}\label{thm:ub_all}
The computation capacity of any sum-network constructed by the SUM-NET-CONS algorithm is at most $1$.
\end{theorem}
\begin{IEEEproof}
By the construction procedure, there is a terminal $t_{p_i}$ which is connected to the sources $s_{p_i}$ and $\{s_{B_j}: A(i,j)=1\}$ through the edge $e_i$. By lemmas \ref{lem:partial_sums} and \ref{lem:partial_sums_indep} we have that $H(\phi_i(X)) \geq m\log_2 q$ bits. From the definition of $n$ the maximum amount of information transmitted on $e_i$ is $n \log_2q$ bits and that implies that $m \leq n$.
\end{IEEEproof}
Next, we show that the upper bound on the computation capacity exhibits a strong dependence on the characteristic of the field (denoted $\ch (\mathcal{F})$) over which the computation takes place.
\begin{theorem}\label{thm:ub}
Let $A$ be a $(0,1)$-matrix of dimension $r \times c$ and suppose that we construct a sum-network corresponding to $A$ using the SUM-NET-CONS algorithm. The matrix $M_A$ is as defined in eq. \eqref{eq:defn_M_A}.
If $\rank_{\mathcal{F}}M_A = r+t$, the upper bound on computation capacity of the sum-network is $r/(r+t)$.
\end{theorem}
\begin{IEEEproof}
Let $\calB_{\mathcal{S}'}\subseteq \calB$ be as defined in lemma \ref{lem:partial_sums_indep}. Then
from lemmas \ref{lem:partial_sums} and \ref{lem:partial_sums_indep}, we have $H\left(X_{p_i}' | \phi_{i}(X)\right)=0,\allowbreak ~\forall i \in [r]$ and $H\left(X_{B_{\mathcal{S}_j'}}'|\{\phi_i(X):A(i,j)=1\}\right)=0,\allowbreak ~\forall j \in [t]$. %\{j: B_{\mathcal{S}_j'} \in \mathcal{B}_{\mathcal{S}'}\}$.
Hence we have that $H(\{\phi_i(X)\}_1^r) \geq m(r+t)\log q$. From the definition of $n$, we get
$nr\log q \geq H(\{\phi_i(X)\}_1^r) \geq m(r+t)\log q \implies m/n \leq r/(r+t)$.
\end{IEEEproof}
\begin{proposition}
  We have that $\rank_{\mathcal{F}}M_A = r+t$ if and only if $\rank_{\mathcal{F}}\left((A^TA)_{\#}-A^TA\right) = t$. Furthermore, $\rank_{\mathcal{F}}M_A=r+c$ if and only if $\ch(\mathcal{F}) \nmid \det_{\mathbb{Z}}M_A$, where $\det_\mathbb{Z}$ indicates the determinant of the matrix with its elements interpreted as $0$ or $1$ in $\mathbb{Z}$.% as opposed to being elements of $\mathcal{F}$.
\end{proposition}
\begin{IEEEproof}
From eq. \eqref{eq:defn_M_A}, we have that
\begin{IEEEeqnarray}{Rl}
M_A&=
\begin{bmatrix}
I_r & A\\
A^T & (A^TA)_{\#}
\end{bmatrix}\IEEEnonumber\\
&=
\begin{bmatrix}
I_r & \bm{0}\\
A^T & I_c
\end{bmatrix}
\begin{bmatrix}
I_r & \bm{0}\\
\bm{0} & (A^TA)_{\#}-A^TA
\end{bmatrix}
\begin{bmatrix}
I_r & A\\
\bm{0} & I_c
\end{bmatrix},\IEEEeqnarraynumspace \label{eq:M_diagonalized}
\end{IEEEeqnarray}
which gives us the rank condition.
Since $M_A$ is a $(0,1)$-matrix, if it has full rank, then its determinant is some non-zero element of $\underline{\mathcal{F}}$, where $\underline{\mathcal{F}}$ is the base subfield of $\mathcal{F}$ having prime order. We could also interpret the elements of $M_A$ as integers and evaluate its determinant $\det_{\mathbb{Z}}M_A$. Then if $M_A$ has full rank, we have that $\ch (\underline{\mathcal{F}}) \nmid \det_{\mathbb{Z}}M_A$.
\end{IEEEproof}
\begin{example}
Consider the normal sum-network obtained from using the Fano plane for which the incidence matrix $A_\mathcal{I}$ is as defined in eq. \eqref{eq:Afano}, so that $r = c =7$. It can be verified that $\rank_{GF(2)}M_{A_\mathcal{I}}=7$. Hence theorem \ref{thm:ub} gives an upper bound of $1$ for the computation capacity. In fact, there is a rate-$1$ network code that satisfies all terminals in the normal sum-network obtained using the Fano plane as described later in proposition \ref{cor:scalar_nw_code}.%Hence the upper bound in Theorem \ref{thm:ub} is not applicable to this sum-network, however Corollary \ref{cor:ub_rank} gives an upper bound of $1$ for the computation capacity. In fact, there is a rate-$1$ network code that satisfies all terminals in the normal sum-network obtained using the Fano plane as described later in Proposition \ref{cor:scalar_nw_code}.
\vspace{2mm}
\end{example}
We can obtain a different upper bound on the computation capacity by considering submatrices of $M_A$ that do not necessarily contain all the initial $r$ rows. To do this we define a new index set $\mathcal{S}^{''}$ based on an index set $\mathcal{S}\subseteq [r]$ as follows.%  For an index set $\mathcal{S}\subseteq [r]$ we define
\begin{IEEEeqnarray}{C}
 \mathcal{S}^{''} \subseteq \{r+1,r+2,\ldots,r+c\} ~\text{such that}\IEEEnonumber\\
  \forall i \in \mathcal{S}^{''}, A^T[i-r,[r]] \in \Span \{I_r[j,[r]]:j \in \mathcal{S}\}.\label{eq:blocks_subset_bottleneck}
\end{IEEEeqnarray}
Here $\Span$ indicates the subspace spanned by the vectors in a set. The submatrix of $M_A$ that contains all the rows indexed by numbers in $\mathcal{S}\cup\mathcal{S}^{''}$ is $M[\mathcal{S}\cup\mathcal{S}^{''},[r+c]]$. %\aditya{Will re-read following theorem.}
\begin{theorem}\label{thm:ub_subset}
  Let $A$ be a $(0,1)$-matrix of dimension $r \times c$ and suppose that we construct a sum-network corresponding to $A$ using the SUM-NET-CONS algorithm. For any $(m,n)$-network code that enables all the terminals to compute the sum, we must have that
  \begin{equation*}
    \frac{m}{n} \leq \min_{\mathcal{S}\subseteq [r]}\left\{\frac{|\mathcal{S}|}{x_\mathcal{S}}\right\},
  \end{equation*} where $x_\mathcal{S}\coloneqq\rank_\mathcal{F}M_A[\mathcal{S}\cup \mathcal{S}^{''},[r+c]]$ and $\mathcal{S}^{''}$ is as defined in eq. \eqref{eq:blocks_subset_bottleneck}.
\end{theorem}
\begin{IEEEproof}
Note that for the choice $\mathcal{S}=[r]$, the index set $\mathcal{S}^{''}$ is the same as the index set $\mathcal{S}'$ defined in lemma \ref{lem:partial_sums_indep} and $x_\mathcal{S}=\rank_\mathcal{F}M_A$, thus recovering the $r/\rank_\mathcal{F}M_A$ upper bound on the computation capacity from theorem \ref{thm:ub}.
For $\mathcal{S}=\{\mathcal{S}_1,\ldots,\mathcal{S}_{|\mathcal{S}|}\}\subset [r]$, we have an index set $\mathcal{T} \subseteq \mathcal{S}^{''}$ such that
\begin{IEEEeqnarray*}{Rl}
 x_{\mathcal{S}} &= \rank_\mathcal{F} M_A[\mathcal{S} \cup \mathcal{S}^{''}, [r+c]],\\
  &= \rank_\mathcal{F} M_A[\mathcal{S} \cup \mathcal{T} ,[r+c]] = |\mathcal{S}|+|\mathcal{T}|.
\end{IEEEeqnarray*}
We collect the blocks indexed in increasing order by $\mathcal{T}$ in the set $\calB_{\mathcal{T}} = \{B_{\mathcal{T}_1},\ldots,B_{\mathcal{T}_y}\} \subseteq \calB$, where $y\coloneqq |\mathcal{T}|$.
Then one can recover the L.H.S. of the following equation from the set of messages $\{\phi_i(X): i \in \mathcal{S}\}$
\begin{IEEEeqnarray*}{C}
  \begin{bmatrix}
  X_{p_{\mathcal{S}_1}}^{'T} & \cdots & X_{p_{\mathcal{S}_{|\mathcal{S}|}}}^{'T} & X_{B_{\mathcal{T}_1}}^{'T} & \cdots & X_{B_{\mathcal{T}_y}}^{'T}
  \end{bmatrix}^T\\
  =\left(\begin{bmatrix}
  M_A[\mathcal{S},[r+c]] \\
  M_A[\mathcal{T},[r+c]]
  \end{bmatrix}\otimes I_m\right) \bm{X}.
\end{IEEEeqnarray*}
Hence we have that $q^{n|\mathcal{S}|}\geq q^{m(|\mathcal{S}|+y)}\implies m/n \leq |\mathcal{S}|/x_\mathcal{S}$. The same reasoning is valid for any choice of $\mathcal{S}\subseteq [r]$ and that gives us the result.
\end{IEEEproof}
\begin{example}\label{eg:disjoint-G}
  \begin{figure}
    \centering
    \includegraphics[width = 0.3 \textwidth]{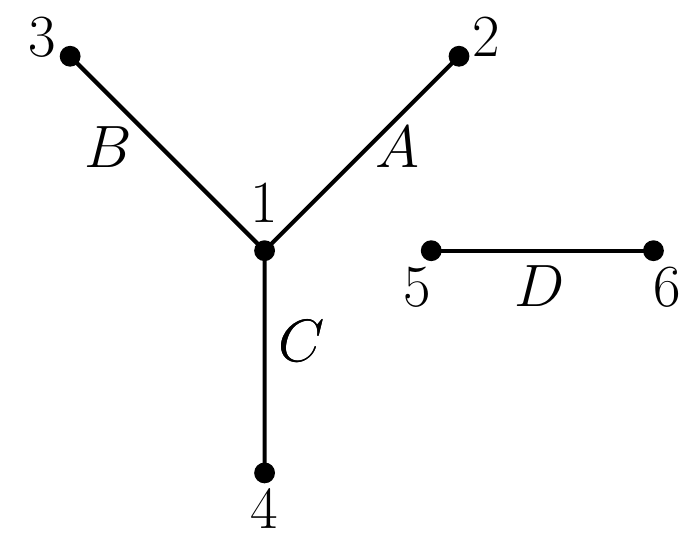}
    \caption{A simple undirected graph $G$ with two connected components. It has $6$ vertices and $4$ edges.}\label{fig:disjoint-G}
  \end{figure}
  Consider the transposed sum-network corresponding to the undirected graph $G$ shown in Figure \ref{fig:disjoint-G}. One can check that the matrix $M_{A_G^T}$ when the rows and columns of the incidence matrix $A_G^T$ are arranged in increasing alphabetical and numeric order is as follows.
  \begin{equation*}
    M_{A_G^T} = \left[\begin{array}{cccc|cccccc}
                  1 & 0 & 0 & 0 & 1 & 1 & 0 & 0 & 0 & 0 \\
                  0 & 1 & 0 & 0 & 1 & 0 & 1 & 0 & 0 & 0 \\
                  0 & 0 & 1 & 0 & 1 & 0 & 0 & 1 & 0 & 0 \\
                  0 & 0 & 0 & 1 & 0 & 0 & 0 & 0 & 1 & 1 \\\hline
                  1 & 1 & 1 & 0 & 1 & 1 & 1 & 1 & 0 & 0 \\
                  1 & 0 & 0 & 0 & 1 & 1 & 0 & 0 & 0 & 0 \\
                  0 & 1 & 0 & 0 & 1 & 0 & 1 & 0 & 0 & 0 \\
                  0 & 0 & 1 & 0 & 1 & 0 & 0 & 1 & 0 & 0 \\
                  0 & 0 & 0 & 1 & 0 & 0 & 0 & 0 & 1 & 1 \\
                  0 & 0 & 0 & 1 & 0 & 0 & 0 & 0 & 1 & 1
                \end{array}\right]
  \end{equation*}
  We choose our finite field alphabet to be $GF(3)$ in this example. Then $\rank_{GF(3)}M_{A_G^T} = 5$ and theorem \ref{thm:ub} gives that the computation capacity is at most $4/5$. However, theorem \ref{thm:ub_subset} gives a tighter upper bound in this case. Specifically, if $\mathcal{S} = \{1,2,3\}$ then $\mathcal{S}^{''}=\{5,6,7,8\}$ and $\rank_{GF(3)}M_{A_G^T}[\mathcal{S}\cup \mathcal{S}^{''},[10]] = 4$. Hence theorem \ref{thm:ub_subset} states that the computation capacity of the transposed sum-network for the graph $G$ is at most $3/4$.
\vspace{2mm}
\end{example}
We apply the above theorems to obtain characteristic dependent upper bounds on the computation capacity of some infinite families of sum-networks constructed using the given procedure.
\begin{corollary}\label{cor:undirected_graph}
Let $\mathcal{I}=(\calP,\mathcal{B})$ be an incidence structure obtained from a simple undirected graph where $\calP$ denotes the set of vertices and $\mathcal{B}$ consists of the $2$-subsets of $\calP$ corresponding to the edges. %Let $\deg(i)\in \mathbb{Z}$ represent the degree of vertex $p_i \in \mathcal{P}$.
Let $\deg(p)\in \mathbb{Z}$ represent the degree of vertex $p \in \mathcal{P}$.
The incidence matrix $A_{\mathcal{I}}$ has dimension $|\calP| \times |\mathcal{B}|$. The computation capacity of the normal sum-network constructed using $A_{\mathcal{I}}$ is at most $\frac{|\calP|}{|\calP|+|\mathcal{B}|}$ for any finite field $\mathcal{F}$.

Let $\mathcal{F}$ be the finite field alphabet of operation %used for communication and define $\calP' \subseteq \calP$ as
 and define $\calP' \subseteq \calP$ as $\calP' \coloneqq \{p: \ch(\mathcal{F})\nmid (\deg(p)-1), p \in \calP\}$.
Consider the set of edges $\calB' \coloneqq\cup_{p \in \calP'} \langle p\rangle $.
%\{B: B \in \langle p \rangle ~\text{for some}~ p \in \calP', B \in \calB\}.
The computation capacity of the transposed sum-network is at most $\frac{|\calB'|}{|\calB'|+|\calP'|}$.
\end{corollary}
\begin{IEEEproof}
Recall that $\bm{B}_i^T$ is the $i$-th row of $A_\mathcal{I}^T$ for all $i \in [|\mathcal{B}|]$. Then the inner product over $\mathcal{F}$ between two rows is
\begin{equation*}
\bm{B}_i^T\bm{B}_j =
\begin{cases}
2\pmod{\ch(\mathcal{F})}, & \text{if~} i = j,\\
1, & \begin{IEEEeqnarraybox*}[][c]{s,} if edges indexed by $i$ and\\$j$ have a common vertex,
                        \end{IEEEeqnarraybox*}\\%\text{~if the edges indexed by $i$ and $j$ have a common vertex,} \\
0, & \text{otherwise}.
\end{cases}
\end{equation*}
It can be observed that the matrix of interest, i.e., $(A_{\mathcal{I}}^TA_{\mathcal{I}})_{\#}-A_{\mathcal{I}}^TA_{\mathcal{I}}=-I_{|\mathcal{B}|}$ has full rank over every finite field.

The transposed sum-network for $\mathcal{I}$ is obtained by applying the SUM-NET-CONS algorithm on the $|\calB|\times |\calP|$ matrix $A^T_\mathcal{I}$, so that the parameters $r = |\calB|, c = |\calP|$. %$\calP' \subseteq \calP$ is the maximal set of points such that it satisfies the required condition.
We apply theorem \ref{thm:ub_subset} by choosing the index set $\mathcal{S}\subseteq [|\calB|]$ such that $\mathcal{S}=\{j: B_j \in \calB'\}$. Defined this way, $|\mathcal{S}|=|\mathcal{B}'|$ and $\mathcal{S}^{''}$ is obtained from $\mathcal{S}$ using eq. \eqref{eq:blocks_subset_bottleneck}. We collect all the points corresponding to the rows in the submatrix $M_{A^T_\mathcal{I}}[\mathcal{S}^{''},[r+c]]$ in a set $\calP_{\mathcal{S}^{''}} \subseteq \calP$.
Note that $\calP_{\mathcal{S}^{''}}$ depends on the set of edges $\calB'$.
By definitions of $\calB'$ and $\mathcal{S}^{''}$, we have that $\calP' \subseteq \calP_{\mathcal{S}^{''}}$. This is true because $\calB'$ consists of all the edges that are incident to at least one point in $\calP'$ while indices in the set $\mathcal{S}^{''}$ correspond to all points that are not incident to any edge outside $\calB'$. For instance, in Example \ref{eg:disjoint-G} above, as $\mathcal{F}=GF(3)$, $\calP' = \{1\}$. Then $\calB' = \{A, B, C\}$ and $\calP_{\mathcal{S}^{''}}=\{1, 2, 3, 4\}$.

We now show that $\rank_\mathcal{F}M_A[\mathcal{S}\cup \mathcal{S}^{''}]=|\calB'|+|\calP'|$ and that gives us the result using theorem \ref{thm:ub_subset}.
Recall that $\bm{p}_i$ denotes the $i$-th row of $A_{\mathcal{I}}$, which corresponds to the vertex $p_i$ for all $i \in [|\calP|]$. It follows that the inner product between $\bm{p}_i, \bm{p}_j$ over $\mathcal{F}$ is
\begin{equation*}
\bm{p}_i \bm{p}_j^T = \begin{cases}
\deg(p_i)\pmod{\ch(\mathcal{F})}, & \text{~if~} i = j,\\
1, & \text{~if~} \{i,j\} \in \mathcal{B},\\
0, & \text{~otherwise}.
\end{cases}
\end{equation*}
Because of the above equation, all the off-diagonal terms in the matrix $(A_\mathcal{I}A_\mathcal{I}^T)_\#-A_\mathcal{I}A_\mathcal{I}^T$ are equal to zero. We focus on the submatrix $M[\mathcal{S}\cup\mathcal{S}^{''},[r+c]]$ obtained from eq. \eqref{eq:M_diagonalized}, letting $\mathcal{S}^{''}_{|\mathcal{B}|}=\{j-|\mathcal{B}|: j \in \mathcal{S}^{''}\}$ we get that
\begin{IEEEeqnarray*}{C}
  M[\mathcal{S}\cup\mathcal{S}^{''},[r+c]]=\begin{bmatrix}
                                   I_{|\calB|}[\mathcal{S},\mathcal{S}] & \bm{0} \\
                                   A_\mathcal{I}[\mathcal{S}^{''}_{|\calB|},\mathcal{S}] & I_{|\calP|}\left[\mathcal{S}^{''}_{|\calB|},\mathcal{S}^{''}_{|\calB|}\right]
                                 \end{bmatrix}\\
                                 \hfill \cdot \Lambda \cdot
                                 \begin{bmatrix}
                                   I_{|\calB|} & A_\mathcal{I}^T \\
                                   \bm{0} & I_{|\calP|}
                                 \end{bmatrix},
\end{IEEEeqnarray*} where
\begin{equation*}
\Lambda \coloneqq
  \begin{bmatrix}
  I_{|\calB|}[\mathcal{S},[|\calB|]] & \bm{0} \\
  \bm{0} & \left((A_\mathcal{I}A_\mathcal{I}^T)_\#-A_\mathcal{I}A_\mathcal{I}^T\right)\left[\mathcal{S}^{''}_{|\calB|},[|\calP|]\right]
  \end{bmatrix}.
\end{equation*}
By definition of $\mathcal{P}'$ the points in the set $\calP_{\mathcal{S}^{''}} \setminus \calP'$ are such that $\deg(p_i)-1 \equiv 0 \pmod{\ch(\mathcal{F})}$, i.e., the diagonal entry corresponding to those points in $(A_\mathcal{I}A_\mathcal{I}^T)_\#-A_\mathcal{I}A_\mathcal{I}^T$ in the matrix $\Lambda$ is zero. Thus, $\Lambda$ has exactly $|\calB'|+|\calP'|$ rows which are not equal to the all-zero row vector. The first and third matrices are invertible, and hence we get that $\rank_\mathcal{F}M_A[\mathcal{S}\cup \mathcal{S}^{''},[r+c]]=|\calB'|+|\calP'|$.
\end{IEEEproof}

\begin{corollary}\label{cor:bibd}
Let $\mathcal{I}=(\calP,\calB)$ be a $2$-$(v,k,1)$ design.
For the normal sum-network constructed using the $|\calP|\times |\calB|$ incidence matrix $A_{\mathcal{I}}$, the computation capacity is at most $\frac{|\calP|}{|\calP|+|\calB|}$ if $\ch (\mathcal{F}) \nmid (k-1)$. For the transposed sum-network constructed using $A_{\mathcal{I}}^T$, the computation capacity is at most $\frac{|\calB|}{|\calP|+|\calB|}$ if $\ch(\mathcal{F}) \nmid \frac{v-k}{k-1}$.
%Consider the normal sum-network constructed using the $|\calP|\times |\calB|$ incidence matrix $A_{\mathcal{I}}$. If $\ch (\mathcal{F}) \nmid (k-1)$, the computation capacity of this sum-network is at most $\frac{|\calP|}{|\calP|+|\calB|}$.
\end{corollary}
\begin{IEEEproof}
We first describe the case of the transposed sum-network. From eq. \eqref{eq:t_design_b_i} each point in a $2$-$(v,k,1)$ design is incident to $\rho= \frac{v-1}{k-1}$ blocks. Moreover any two points occur together in exactly one block. Thus, we have the inner product over $\mathcal{F}$ as
\begin{equation*}
\bm{p}_i\bm{p}_j^T=
\begin{cases}
  \frac{v-1}{k-1}\pmod{\ch(\mathcal{F})}, & \mbox{if } j=i, \\
  1, & \mbox{otherwise}.
\end{cases}
\end{equation*}
This implies that $A_{\mathcal{I}}A_{\mathcal{I}}^T-(A_{\mathcal{I}}A_{\mathcal{I}}^T)_{\#}=\left[\left(\frac{v-1}{k-1}-1\right)\right]I_{v}=\left[\frac{v-k}{k-1}\right]I_v$ and setting its determinant non-zero gives the result.

For the normal sum-network, we argue as follows.
Note that $\bm{B}_i^T\bm{B}_i=k\pmod{\ch(\mathcal{F})}$ for any $i$. Since any two points determine a unique block, two blocks can either have one point or none in common. Hence, for $i \neq j$, the inner product over $\mathcal{F}$ is
\begin{equation*}
\bm{B}_i^T\bm{B}_j =
\begin{cases}
1, & \text{~if~} B_i \cap B_j \neq \emptyset,\\
0, & \text{~otherwise}.
\end{cases}
\end{equation*}

Then $A_{\mathcal{I}}^TA_{\mathcal{I}}-(A_{\mathcal{I}}^TA_{\mathcal{I}})_{\#}=\left[(k-1)\right]I_b$ and setting its determinant as non-zero gives the result.
\end{IEEEproof}
\begin{corollary}
Let $\mathcal{I}=(\calP, \calB)$ be a $t$-$(v,k,\lambda)$ design, for $t \geq 2$. From eq. \eqref{eq:t_design_b_i},  each point is present in $
\rho \mathrel{\mathop:}= \lambda\binom{v-1}{t-1}/\binom{k-1}{t-1}
$ blocks and the number of blocks incident to any pair of points is given by $
b_2 \mathrel{\mathop:}= \lambda\binom{v-2}{t-2}/\binom{k-2}{t-2}
$.
Consider the transposed sum-network constructed using the incidence matrix $A_{\mathcal{I}}^T$ which has dimension $|\calB|\times |\calP|$. The computation capacity of the transposed sum-network is at most $\frac{|\calB|}{|\calB|+|\calP|}$ if
\begin{equation*}
\ch (\mathcal{F})~\nmid~ [\rho-b_2+v(b_2-1)](\rho-b_2)^{v-1}.
\end{equation*}
\end{corollary}
\begin{IEEEproof}
%Let $\bm{p}_i$ denote the $i$-th row of $A_{\mathcal{I}}$, which corresponds to a point $p_i \in \calP$ for all $i \in [v]$.
By definition, we have that the inner product over $\mathcal{F}$ between two rows is
\begin{equation*}
\bm{p}_i\bm{p}_j^T=
\begin{cases}
  \rho \pmod{\ch(\mathcal{F})}, & \mbox{if } j=i, \\
  b_2 \pmod{\ch(\mathcal{F})}, & \mbox{otherwise}.
\end{cases}
\end{equation*}
%$\bm{p}_i\bm{p}_j^T=\rho\mod\ch(\mathcal{F})$ if $j = i$ and $b_2\mod\ch(\mathcal{F})$ otherwise.
It follows that $A_{\mathcal{I}}A_{\mathcal{I}}^T-(A_{\mathcal{I}}A_{\mathcal{I}}^T)_{\#}$ has the value $(\rho-1)$ on the diagonal and $(b_2-1)$ elsewhere. That is,
\begin{IEEEeqnarray*}{L}
A_{\mathcal{I}}A_{\mathcal{I}}^T-(A_{\mathcal{I}}A_{\mathcal{I}}^T)_{\#}\\
= \left[(\rho\!-\!b_2)\!\!\!\pmod{\ch(\mathcal{F})}\right]I_{v}+\left[(b_2\!-\!1)\!\!\!\pmod{\ch(\mathcal{F})}\right]J_{v},
\end{IEEEeqnarray*}
where $J_{v}$ denotes the square all ones matrix of dimension $v$.
Then by elementary row and columns operations, $\det\left[A_{\mathcal{I}}A_{\mathcal{I}}^T-(A_{\mathcal{I}}A_{\mathcal{I}}^T)_{\#}\right]$ can be evaluated to be equal to $[\rho-b_2+v(b_2-1)](\rho-b_2)^{v-1}\pmod{\ch(\mathcal{F})}$.
\end{IEEEproof}

\begin{corollary}\label{cor:tdesign_higher_inc}
Let $\calD = (\calP, \calB)$ be a $t$-$(v,t+1,\lambda)$ design with $\lambda \neq 1$ and incidence matrix $A_\calD$. We define a \textit{
higher} incidence matrix $A_{\calD'}$ of dimension $\binom{|\calP|}{t} \times |\calB|$ such that each row corresponds to a distinct $t$-subset of $\calP$ and each column corresponds to a block in $\calB$. $A_{\calD'}$ is a $(0,1)$-matrix such that for any $i \in \left[\binom{v}{t}\right], j \in [|\calB|]$, its entry $A_{\calD'}(i,j)=1$ if each of the points in the $t$-subset corresponding to the $i$-th row is incident to the block $B_j \in \calB$ and zero otherwise. The computation capacity of the normal sum-network constructed using $A_{\calD'}$ is at most $\frac{\binom{v}{t}}{\binom{v}{t} + |\calB|} = \frac{t+1}{\lambda + t + 1}$ if $\ch(\mathcal{F}) \nmid t$. The computation capacity of the transposed sum-network constructed using $A_{\calD'}^T$ is at most $\frac{|\calB|}{|\calB| + \binom{v}{t}}=\frac{\lambda}{\lambda+t+1}$ if $\ch(\mathcal{F}) \nmid (\lambda - 1)$.
%Using $\calD$, we define a \textit{higher} incidence structure $\mathcal{I}=(\calP,\calB)$ whose point set $\calP$ consists of \textit{all} $t$-subsets of the point set $\calP_\calD$. The block set $\calB$ is set to equal $\calB_\calD$. A block is incident to a point $p \in \calP$ if it contains all the $t$ points from $\calP_\calD$ that are present in $p$.
\end{corollary}
\begin{IEEEproof}
The incidence matrix $A_{\calD'}$ is a $(0,1)$ matrix of dimension $\binom{v}{t} \times \frac{\lambda}{t+1}\binom{v}{t}$.
Let $\bm{p}_i, \bm{B}_u$ denote the $i$-th row and $u$-th column respectively of $A_\mathcal{D'}$ for $i \in \left[\binom{v}{t}\right], u \in \left[\frac{\lambda}{t+1}\binom{v}{t}\right]$. Each row of $A_{\mathcal{D}'}$ corresponds to a distinct $t$-subset of $\calP$. By $t$-design criterion, any set of $t$ points belongs to exactly $\lambda$ blocks. Since the columns have a one-to-one correspondence with the blocks in $\calB$, each row of $A_{\mathcal{D}'}$ has exactly $\lambda$ $1$'s. Two rows will have a $1$ in the same column if the block corresponding to the column is incident to both the $t$-subsets corresponding to the two rows. Since each block has $t+1$ points, there cannot be more than one block incident to two different $t$-subsets. Hence, for the inner product over $\mathcal{F}$, we have that $\bm{p}_i\bm{p}_i^T=\lambda\pmod{\ch(\mathcal{F})}$ and for all $i \neq j; i,j, \in \left[\binom{v}{t}\right]$,
\begin{equation*}
\bm{p}_i\bm{p}_j^T=
\begin{cases}
1, & \begin{IEEEeqnarraybox*}[][c]{!s,} if the union of the $t$-subsets corresponding to\\the $i$-th and $j$-th rows is a block in $\calB$,
 \end{IEEEeqnarraybox*}\\%\text{~if the edges indexed by $i$ and $j$ have a common vertex,} \\
0, & \!\text{otherwise}.
\end{cases}
\end{equation*}
Then $A_\mathcal{D'}A_{\mathcal{D'}}^T-(A_\mathcal{D'}A_{\mathcal{D'}}^T)_{\#}=\left[(\lambda -1)\pmod{\ch(\mathcal{F})}\right]I_{\binom{v}{t}}$ and that gives the result for the transposed sum-network.

For the normal sum-network, we look at the columns of $A_{\mathcal{D}'}$ in a similar manner. Each column of $A_{\mathcal{D}'}$ corresponds to a block in $\calB$. Since the size of each block is $t+1$, each column has exactly $\binom{t+1}{t}=t+1$ elements as $1$. Also, two different blocks can have at most $t$ points in common, and only when that happens, will the two columns have a $1$ in the same row. Hence, for the inner product over $\mathcal{F}$, we have that $\bm{B}_u^T\bm{B}_u=(t+1)\pmod{\ch(\mathcal{F})}$ and for all $u \neq v; u,v \in \left[\binom{v}{t}\right]$,
\begin{equation*}
\bm{B}_u^T\bm{B}_v=
\begin{cases}
1, & \begin{IEEEeqnarraybox*}[][c]{s,} if the $u$-th and $v$-th blocks have $t$ points\\in common,
 \end{IEEEeqnarraybox*}\\%\text{~if the edges indexed by $i$ and $j$ have a common vertex,} \\
0, & \text{otherwise}.
\end{cases}
\end{equation*}
Then $A_\mathcal{D'}^TA_{\mathcal{D}'}-(A_\mathcal{D'}^TA_{\mathcal{D'}})_{\#}=t\pmod{\ch(\mathcal{F})}I_{\frac{\lambda}{t+1}\binom{v}{t}}$ and theorem \ref{thm:ub} gives the result. % Thus  in both cases and we obtain the result.
\end{IEEEproof}

\section{Linear network codes for constructed sum-networks}\label{sec:lower_bd}
In this section, we propose linear network codes for the sum-networks constructed using the \sumnetalg algorithm. Recall that the algorithm takes a $(0,1)$-matrix $A$ that has $r$ rows and $c$ columns as its input. In Section \ref{sec:upper_bd}, we demonstrated that the incidence matrix of certain incidence structures result in sum-networks whose capacity can be upper bounded ({\it cf.} Corollaries \ref{cor:undirected_graph}, \ref{cor:bibd}, \ref{cor:tdesign_higher_inc}). We now demonstrate that under certain conditions, we can obtain network codes whose rate matches the corresponding upper bound. Thus, we are able to characterize the capacity of a large family of sum-networks.

We emphasize that random linear network codes that have been used widely in the literature for multicast code constructions are not applicable in our context. In particular, it is not too hard to argue that a random linear network code would result in each terminal obtaining a different linear function or subspace. Thus, constructing codes for these sum-networks requires newer ideas. We outline the key ideas by means of the following example.

\begin{example}\label{eg:K2_code}
Consider the sum-network shown in Figure \ref{fig:K2normal}. The matrix $A_{\mathcal{I}}$ used in its construction is of dimension $r \times c$ where $r=2, c = 1$ and is described in Example \ref{eg:K2}. It can be observed that $A_{\mathcal{I}}^TA_{\mathcal{I}} - \left(A_{\mathcal{I}}^TA_{\mathcal{I}}\right)_\#=1$. Then theorem \ref{thm:ub} states that the computation capacity of this sum-network is at most $2/3$. We describe a network code with $m=2, n=3$. The global encoding functions for the two bottleneck edges are shown in Table \ref{tab:K2_code}.
\begin{table}[h]
\caption{The function values transmitted across $e_1, e_2$ in Figure \ref{fig:K2normal} for a network code with rate $=2/3$. Each message $X_1,\allowbreak X_2,\allowbreak X_{\{1,2\}}$ is a vector with $2$ components, and $\phi_1(X),\allowbreak \phi_2(X)$ are vectors with $3$ components each. A number within square brackets adjoining a vector indicates a particular component of the vector.}
\label{tab:K2_code}
\centering
\begin{tabular}{ccc}
\hline\hline
Component & $\phi_1(X)$ & $\phi_2(X)$ \\ [0.5ex]
\hline
$1$ & $X_1[1]+X_{\{1,2\}}[1]$ & $X_2[1]+X_{\{1,2\}}[1]$\\
$2$ & $X_1[2]+X_{\{1,2\}}[2]$ & $X_2[2]+X_{\{1,2\}}[2]$\\
$3$ & $X_{\{1,2\}}[1]$ & $X_{\{1,2\}}[2]$\\ [1ex]
\hline
\end{tabular}
\end{table}
Using the values transmitted, all three terminals can recover the sum in the following manner. $t_1$ receives the value of $X_2$ from the direct edge $(s_2,t_1)$ while $t_2$ receives the value of $X_1$ from the direct edge $(s_1,t_2)$. Then $t_1$ recovers the sum using the first two components of $\phi_1(X)$ while $t_2$ recovers the sum using the first two components of $\phi_2(X)$. Additionally, $t_{\{1,2\}}$ receives both $\phi_1(X), \phi_2(X)$ and can carry out the operation $(X_1+X_{\{1,2\}}) + (X_2+X_{\{1,2\}}) - X_{\{1,2\}}$. Thus, each terminal is satisfied.
\vspace{2mm}
\end{example}

The network code in the example has the following structure. For each bottleneck edge, the first $r$ components of the global encoding vector are the sum of all messages that are incident to that bottleneck. The remaining $c$ components of the encoding vectors transmit certain components of messages observed at source nodes that correspond to columns in the matrix $A_\mathcal{I}$. In the example, $t_{\{1,2\}}$ received the first component of $X_{\{1,2\}}$ from $\phi_1(X)$ and the second component from $\phi_2(X)$. Thus it was able to recover the value of $X_{\{1,2\}}$, which it used in computing the demanded sum.

Our construction of network codes for sum-networks will have this structure, i.e., the first $r$ components on a bottleneck edge will be used to transmit a \textit{partial} sum of the messages observed at the sources that are connected to that bottleneck edge and the remaining $c$ components will transmit portions of certain sources in an uncoded manner. For a given incidence matrix $A$, our first step is to identify (if possible) a corresponding non-negative integral matrix $D$ of the same dimensions with the following properties.
\begin{itemize}
\item $D(i,j) = 0$ if $A(i,j) = 0$.
\item Each row in $D$ sums to $r$.
\item Each column in $D$ sums to $c$.
\end{itemize}
Under certain conditions on the incidence matrix $A$, we will show that $D$ can be used to construct suitable network codes for the sum-networks under consideration.

The existence of our proposed network codes are thus intimately related to the existence of non-negative integral matrices that satisfy certain constraints. The following theorem \cite[Corollary 1.4.2]{CMC_Brualdi} is a special case of a more general theorem in \cite{mirsky68} that gives the necessary and sufficient conditions for the existence of non-negative integral matrices with constraints on their row and column sums. We give the proof here since we use some ideas from it in the eventual network code assignment. %completeness. %Stated in this manner, it can be proved using the feasible flow theorem for networks with multiple sources and sinks \cite{gale57}.
\begin{theorem}\label{thm:matrix}
Let $R=(r_1,\allowbreak r_2,\allowbreak \ldots,\allowbreak r_m)$ and $S=(s_1,\allowbreak s_2,\allowbreak \ldots,\allowbreak s_n)$ be non-negative integral vectors satisfying $r_1+\ldots+r_m=s_1+\ldots+s_n$. There exists an $m \times n$ nonnegative integral matrix $D$ such that %whose elements and row and column sums satisfy
\begin{IEEEeqnarray*}{Rl}
0 \leq D(i,j) \leq & c_{ij}, \;\; \forall i \in [m], \forall j \in [n],\\
\sum_{j=1}^{n} D(i,j) = & r_i, \;\; \forall i \in [m], \text{~and}\\
\sum_{i=1}^{m} D(i,j) = & s_j, \;\; \forall j \in [n]
\end{IEEEeqnarray*}
if and only if for all $I \subseteq [m]$ and $J \subseteq [n]$, we have that
\begin{equation}\label{eq:thm_matrix}
\sum_{i \in I}\sum_{j \in J}c_{ij} \geq \sum_{j \in J}s_j - \sum_{i \notin I}r_i.
\end{equation}
\end{theorem}
\begin{IEEEproof}
Consider a capacity-limited flow-network modelled using a bipartite graph on $m+n$ nodes. The left part has $m$ nodes denoted as $x_i, \forall i \in [m]$ and the right part has $n$ nodes denoted as $y_j, \forall j \in [n]$. For all $i,j$ there is a directed edge $(x_i,y_j)$ with capacity $c_{ij}$. There are two additional nodes in the flow-network, the source node $S^*$ and terminal node $T^*$. There are directed edges $(S^*,x_i)$ with capacity $r_i$ for all $i \in [m]$ and directed edges $(y_j,T^*)$ with capacity $s_j$ for all $j \in [n]$. Let $x_I$ be the set of all nodes in the left part whose indices are in $I$ and let $y_{\bar{J}}$ be the set of all nodes in the right part whose indices are \textit{not} in $J$. Consider a cut separating nodes in $\{S^*\}\cup x_I \cup y_{\bar{J}}$ from its complement. Let $f^*$ be the value of the maximum $S^*$-$T^*$ flow in this network. Then we must have that for all possible choice of subsets $I \subseteq [m], J \subseteq [n]$,
\begin{equation}\label{eq:thm_matrix_flow}
  \sum_{i \notin I} r_i + \sum_{(i,j): i \in I, j \in J} c_{ij} + \sum_{j \notin J}s_j \geq f^*.
\end{equation}
In particular, suppose that $f^* = \sum_{j \in [n]}s_j$ in the flow-network. Substituting this in eq. \eqref{eq:thm_matrix_flow}, we get the condition that for all possible subsets $I \subseteq [m], J \subseteq [n]$,
\begin{equation}\label{eq:thm_matrix_nec}
  \sum_{i \in I}\sum_{j \in J}c_{ij} \geq \sum_{j \in J}s_j - \sum_{i \notin I}r_i.
\end{equation}
Note that by choosing all possible subsets $I,J$, we are considering every possible $S^*$-$T^*$ cut in the network. Then by the maxflow-mincut theorem, the set of conditions of the form of eq. \eqref{eq:thm_matrix_nec} for all $I,J$ are not only necessary but also sufficient for the existence of a flow of value $f^*=\sum_{j \in [n]}s_j$ in the network.

A feasible flow with this value can be used to arrive at the matrix $D$ as follows. We set the value of element $D(i,j)$ in the matrix to be equal to the value of the feasible flow on the edge $(x_i,y_j)$ for all $i \in [m], j \in [n]$. It is easy to verify that the matrix $D$ satisfies the required conditions.
\end{IEEEproof}
Using the existence theorem for nonnegative integral matrices, we can obtain network codes for sum-networks constructed from certain incidence structures. The following theorem describes a set of sufficient conditions that, if satisfied by an incidence structure, allow us to construct a linear network code that has the same rate as the computation capacity of that sum-network. The proof of the theorem is constructive and results in an explicit network code.
\begin{theorem}\label{thm:lb}
Let $\mathcal{I}=(\calP,\mathcal{B})$ be an incidence structure and let $A_{\mathcal{I}}$ denote the corresponding incidence matrix of dimension $v \times b$. Suppose that the following conditions are satisfied.
\begin{itemize}
\item $A_{\mathcal{I}}^TA_{\mathcal{I}} - (A_{\mathcal{I}}^TA_{\mathcal{I}})_{\#}=\diag(\mu_1,\ldots,\mu_{b}) \pmod{ \ch(\mathcal{F})}$, where $\mu_i$ is a non-zero element of $\mathcal{F} ~\forall i\in \{1,2,\ldots,b\}$.
\item There exists a matrix $D_{\mathcal{I}}$ with integer elements of the same dimension as $A_\mathcal{I}$ whose entries satisfy
\begin{align}
D_\mathcal{I}(i,j) &= 0, \text{~if~} A_\calI(i,j) = 0, \label{eq:D1}\\
\sum_{i=1}^{v}D_\mathcal{I}(i,j) &=v, \text{~and} \label{eq:D2}\\
\sum_{j=1}^{b}D_\mathcal{I}(i,j) &=b. \label{eq:D3}
\end{align}

\end{itemize}
Then the computation capacity of the sum-network constructed using $A_{\mathcal{I}}$ via the \sumnetalg algorithm is $\frac{v}{v+b}$. This rate can be achieved by a linear network code.

\end{theorem}
\begin{IEEEproof}
%Consider a sum-network obtained by applying the construction procedure on the matrix $A_{\mathcal{I}}$.
Note that $A_{\mathcal{I}}^TA_{\mathcal{I}} - (A_{\mathcal{I}}^TA_{\mathcal{I}})_{\#}$ has full rank by assumption, theorem \ref{thm:ub} states that the computation capacity of the sum-network is at most $v/(v+b)$. We construct a $(m,n)$ linear network code with $m=v, n=v+b$ using the matrix $D_{\mathcal{I}}$. Since $m=v$, each message vector has $v$ components. For a vector $t \in \mathcal{F}^v$, the notation $t[l_1:l_2]$ for two positive integers $l_1,l_2 \in [v]$ denotes a $(l_2-l_1+1)$ length vector that contains the components of $t$ with indices in the set $\{l_1,l_1+1,\ldots, l_2\}$ in order. We need to specify the global encoding vectors $\phi_i(X)$ only for the bottleneck edges $e_i, i \in [v]$ as all the other edges in the network act as repeaters. The linear network code is such that the first $v$ components of the vector transmitted along $e_i ~\forall i \in [v]$ is %\aditya{has the notation used below been introduced before?}
\begin{equation*}
  \phi_i(X)[1:v]=X_{p_i} + \sum_{j:A_\mathcal{I}(i,j)=1}X_{B_j}.
\end{equation*}
By construction, each $t_{p_i} \forall i \in [v]$ is connected to the source nodes in $\{s_{p_{i'}}:i' \neq i\} \cup \{s_{B_j}: A_\mathcal{I}(i,j)=0\}$ by direct edges. $t_{p_i}$ can then compute the following value from the information received on the direct edges.
\begin{equation*}
  \sum_{i' \neq i}X_{p_i} + \sum_{j:A_\mathcal{I}(i,j)=0} X_{B_j}.
\end{equation*}
Adding the above value to $\phi_i(X)[1:v]$ enables $t_{p_i}$ to compute the required sum. In what follows, we focus on terminals of the form $t_{B_j}\forall j \in [b]$.

Since $n=v+b$, each vector $\phi_i(X)\in \mathcal{F}^n$ has $b$ components that haven't been specified yet. %In a structured linear network code, these $b$ components contain certain components of some messages from the set $\{X_{B_j}:A_\mathcal{I}(i,j)=1\}$.
We describe a particular assignment for the $b$ components on every $\phi_i(X), i \in [v]$ using the matrix $D_\mathcal{I}$ that enables each $t_{B_j} \forall j \in [b]$ to compute the sum.

Recall the bipartite flow network constructed in the proof of theorem \ref{thm:matrix}. The nodes in the left part are denoted as $p_i \forall i \in [v]$ and the nodes in the right part are denoted as $B_j \forall j \in [b]$. There is an edge $(p_i,B_j)$ if and only if $A_\mathcal{I}(i,j)=1$. The flow on the edge $(p_i,B_j)$ is denoted as $f(p_i,B_j)$ and its value is determined by $D_\mathcal{I}(i,j)$, i.e.,
$
  f(p_i,B_j)\coloneqq D_\mathcal{I}(i,j).
$

By constraints on the row and column sums of $D_\mathcal{I}$, we conclude that the value of the flow through any $p_i \forall i \in [v]$ is $b$ and the value of the flow through any $B_j \forall j \in [b]$ is $v$. Without loss of generality, assume that $B_j = \{p_1,p_2,\cdots, p_{|B_j|}\}$. We can partition the $v$ components of message vector $X_{B_j}$ into $|B_j|$ parts such that the $i$-th partition contains $f(p_i,B_j)$ distinct components of $X_{B_j}$. Such a partitioning can be done for all message vectors $X_{B_j}, j \in [b]$. Then the flow $f(p_i,B_j)$ indicates that the vector $\phi_i(X)[v+1:v+b]$ includes $f(p_i,B_j)$ uncoded components of $X_{B_j}$. Assigning such an interpretation to every edge in the flow-network is possible as the total number of components available in each $\phi_i(X)$ is $b$ and that is also equal to the flow through the point $p_i$.

By construction, terminal $t_{B_j}$ is connected to all bottleneck edges in the set $\{e_i : A_\mathcal{I}(i,j)=1\}$. From the assignment based on the flow, $t_{B_j}$ receives $f(p_i,B_j)$ distinct components of $X_{B_j}$ from $\phi_i(X)$ for all $\{i:A_\mathcal{I}(i,j)=1\}$. Since $\sum_{i=1}^{v}f(p_i,B_j)=v$, it can recover all $v$ components of $X_{B_j}$ in a piecewise fashion.

By adding the first $v$ components transmitted on all the bottleneck edges that are connected to $t_{B_j}$, it can recover
\begin{IEEEeqnarray*}{L}
\sum_{i:A_\mathcal{I}(i,j)=1}\phi_i(X)[1:v] \\
= \sum_{i: A_\mathcal{I}(i,j)=1}X_{p_i} + \sum_{i: A_\mathcal{I}(i,j)=1}\sum_{l:A_\mathcal{I}(i,l)=1}X_{B_l},\\
= \sum_{i: A_\mathcal{I}(i,j)=1}X_{p_i} + \sum_{B_l \in \langle B_j\rangle}\bm{B}_j^T\bm{B}_l X_{B_l}.
\end{IEEEeqnarray*}

Because of the condition that $A_{\mathcal{I}}^TA_{\mathcal{I}} - (A_{\mathcal{I}}^TA_{\mathcal{I}})_{\#}=\diag(\mu_1,\allowbreak \mu_2,\allowbreak \ldots,\allowbreak \mu_b)$, one can verify that
\begin{equation*}
\sum_{B_l \in \langle B_j\rangle}\bm{B}_j^T\bm{B}_l X_{B_l} = (\mu_j + 1)X_{B_j} + \sum_{B_l \in \langle B_j\rangle \setminus B_j} X_{B_l}.
\end{equation*}
By the flow-based assignment, each $t_{B_j}$ obtains the value of $X_{B_j}$ in a piecewise manner. It can then carry out the following
\begin{IEEEeqnarray*}{L}
\sum_{i:A_\mathcal{I}(i,j)=1}\phi_i(X)[1:v] - \mu_jX_{B_j}\\
= \!\!\sum_{i: A_\mathcal{I}(i,j)=1}\!\!X_{p_i} + (\mu_j + 1)X_{B_j} +\!\! \sum_{B_l \in \langle B_j\rangle \setminus B_j} \!\!X_{B_l}- \mu_jX_{B_j},\\
= \sum_{p \in B_j}X_{p} + \sum_{B_l \in \langle B_j\rangle} X_{B_l}.
\end{IEEEeqnarray*}
The messages not present in this partial sum, i.e., $\{X_p: p \notin B_j\} \cup \{X_B : B \notin \langle B_j \rangle\}$ are available at $t_{B_j}$ through direct edges by construction. Hence, terminals that correspond to a column of $A_{\mathcal{I}}$ are also able to compute the required sum.
\end{IEEEproof}
%Theorem \ref{thm:ub} gives an upper bound on the computation capacity of sum-networks constructed using certain incidence structures while theorem \ref{thm:lb} provides a structured linear network code that achieves the capacity for incidence structures that satisfy the qualifying conditions.

We illustrate the linear network code proposed above by means of the following example.

\begin{figure}
\begin{center}
\subfigure[]{\includegraphics[scale=0.65]{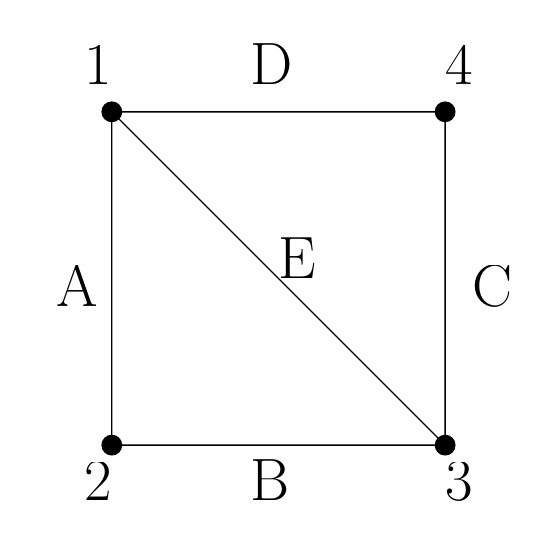}
\label{fig:G}}
\subfigure[]{\includegraphics[scale=0.6]{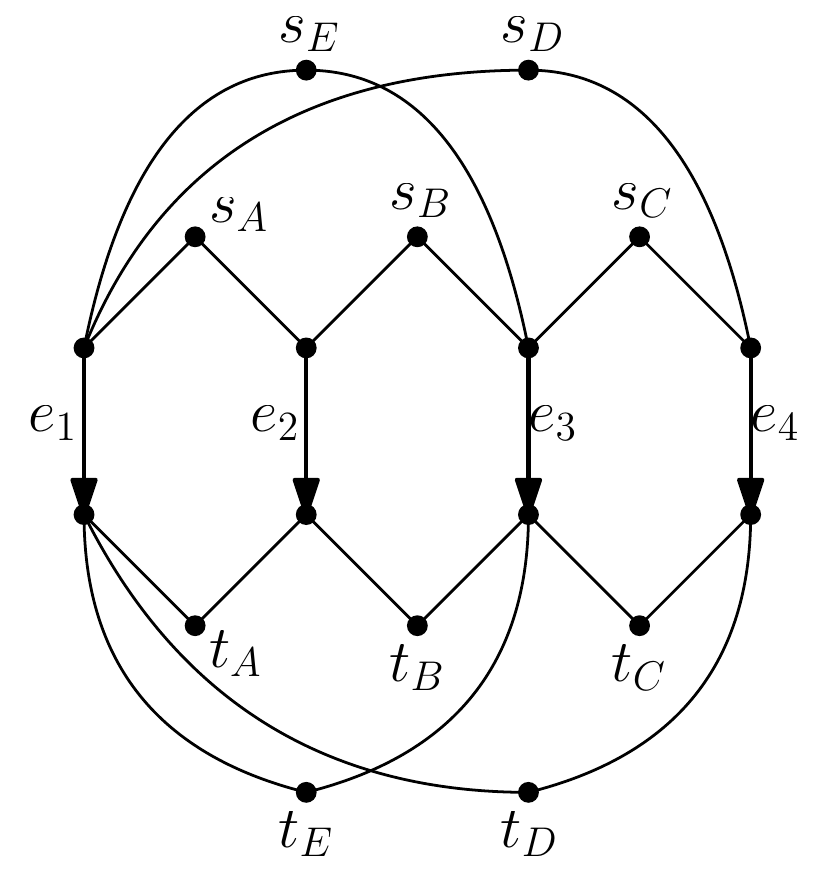}
\label{fig:Gnormal}}
\subfigure[]{\includegraphics[scale = 0.65]{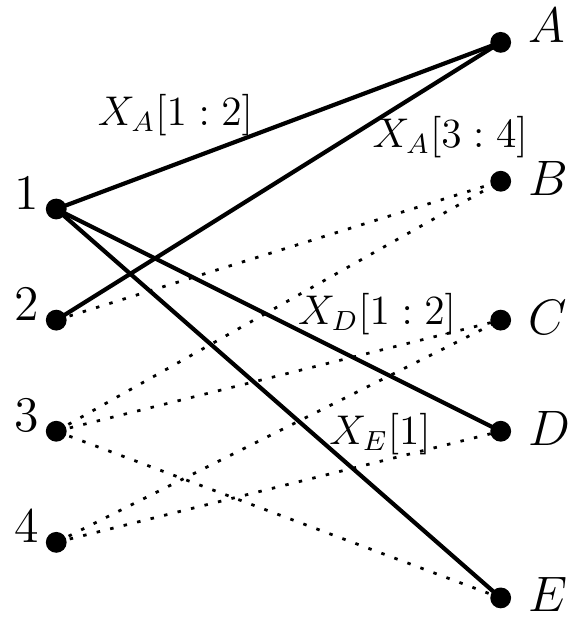}
\label{fig:flow-nw}}
\caption{(a) Undirected graph considered in Example \ref{eg:G}. (b) Part of the corresponding normal sum-network constructed for the undirected graph in (a). The full normal sum-network has nine nodes each in the source set $S$ and the terminal set $T$. However, for clarity, only the five sources and terminals that correspond to the columns of the incidence matrix of the graph are shown. Also, the direct edges constructed in Step 4 of the construction procedure are not shown. All edges are unit-capacity and point downward. The edges with the arrowheads are the bottleneck edges constructed in step 2 of the construction procedure. (c) Bipartite flow network as constructed in the proof of theorem \ref{thm:matrix} for this sum-network. The message values corresponding to the flow on the solid lines are also shown. }
\end{center}
\end{figure}

\begin{example}\label{eg:G}
%\aditya{in above figure 4c, why are some edges solid and others dotted?}
Consider the normal sum-network obtained from the undirected simple graph $G$ shown in Figure \ref{fig:G}. A part of the sum-network is shown in Figure \ref{fig:Gnormal}. The $4 \times 5$ incidence matrix $A_G$ satisfies the condition of theorem \ref{thm:matrix} and therefore has an associated matrix $D_G$ with row-sum as $5$ and column-sum $4$ as shown below. The rows and columns of $A_G$ are arranged in increasing numeric and alphabetical order.
\begin{equation*}
A_G=\begin{bmatrix}
1 & 0 & 0 & 1 & 1\\
1 & 1 & 0 & 0 & 0\\
0 & 1 & 1 & 0 & 1\\
0 & 0 & 1 & 1 & 0
\end{bmatrix},~~
D_G=\begin{bmatrix}
2 & 0 & 0 & 2 & 1\\
2 & 3 & 0 & 0 & 0\\
0 & 1 & 1 & 0 & 3\\
0 & 0 & 3 & 2 & 0
\end{bmatrix}.
\end{equation*}
Using the matrix $D_G$, one can construct a structured linear network code with rate $=v/(v+b)=4/9$ as shown in Table \ref{tab:G_code}. One can check that it enables all the terminals to compute the required sum. The flow-network corresponding to $D_G$ is shown in Figure \ref{fig:flow-nw}, and the messages corresponding to the flow on the solid edges are shown alongside the respective edge.

%\aditya{discuss para with AT}
We can also consider the transposed sum-network for the same graph $G$. Corollary \ref{cor:undirected_graph} gives an upper bound on the computation capacity that depends on $\mathcal{F}$. If $\mathcal{F}=GF(2)$, then the subset of points $\mathcal{P}'=\{2,4\}$ and the upper bound is $4/6$. Note that theorem \ref{thm:lb} is not applicable here as the matrix $A_{G}^TA_{G} - (A_{G}^TA_{G})_{\#}$ does not have all its diagonal elements as non-zero over $GF(2)$. Proposition \ref{claim:irr-graph_lb} gives a condition for the existence of a network code for transposed sum-networks obtained using irregular graphs. We apply that condition to the transposed sum-network of the graph $G$ considered here in Example \ref{eg:irr-graph_transpose}.%\aditya{Need a figure that shows the labeling of some edges of the bipartite flow network}
\begin{table*}
\caption{The function values transmitted across $e_1, e_2, e_3, e_4$ in Figure \ref{fig:Gnormal} for a network code with rate $=4/9$. Each message $X_A,X_B,X_C,X_D,X_E$ is a vector with $4$ components, and $\phi_1(X),\phi_2(X), \phi_3(X), \phi_4(X)$ are vectors with $9$ components each. The number inside square brackets adjoining a vector indicates a particular component of the vector.}
\label{tab:G_code}
\centering
\begin{tabular}{ccccc}
\hline\hline
Component & $\phi_1(X)$ & $\phi_2(X)$ & $\phi_3(X)$ & $\phi_4(X)$\\ [0.5ex]
\hline
$1$ to $4$ & $X_1+X_A+X_D+X_E$ & $X_2+X_A+X_B$ & $X_3+X_B+X_C+X_E$ & $X_4+X_C+X_D$\\
$5$ & $X_A[1]$ & $X_A[3]$ & $X_B[4]$ & $X_C[2]$\\
$6$ & $X_A[2]$ & $X_A[4]$ & $X_C[1]$ & $X_C[3]$\\
$7$ & $X_D[1]$ & $X_B[1]$ & $X_E[2]$ & $X_C[4]$\\
$8$ & $X_D[2]$ & $X_B[2]$ & $X_E[3]$ & $X_D[3]$\\
$9$ & $X_E[1]$ & $X_B[3]$ & $X_E[4]$ & $X_D[4]$\\
[1ex]
\hline
\end{tabular}
\end{table*}
\vspace{2mm}
\end{example}
In the following proposition we show that certain infinite families of incidence structures satisfy the requirements stated in theorem \ref{thm:lb}. In particular, the incidence structures considered in Corollaries \ref{cor:undirected_graph}, \ref{cor:bibd} and \ref{cor:tdesign_higher_inc} satisfy the conditions and hence the computation capacity of the associated sum-networks can be calculated.

\begin{proposition}\label{thm:capacity}
The following incidence structures and their transposes satisfy condition (ii) in theorem \ref{thm:lb}, i.e., if their incidence matrix of dimension $v \times b$ is denoted by $A_\calI$, there exists a corresponding non-negative integral matrix $D_\calI$ that satisfies the conditions in equations (\ref{eq:D1}) -- (\ref{eq:D3}).
\begin{enumerate}
\item Incidence structures derived from a regular graph or a biregular bipartite graph.
\item $t$-$(v,k,\lambda)$ designs with $\lambda = 1$.
\item The higher incidence structure of a $t$-$(n,t+1,\lambda)$ design with $\lambda \neq 1$ obtained using the procedure described in corollary \ref{cor:tdesign_higher_inc}.
\end{enumerate}
\end{proposition}

\begin{IEEEproof}
The existence of $D_{\mathcal{I}}$ with row-sums as $v$ and column-sums $b$ is the same as the existence of $D^{T}_{\mathcal{I}}$ with row-sums as $b$ and column-sums $v$. Thus, it suffices to argue for $D_{\mathcal{I}}$.
To check the validity of the condition we first choose the bounds on the elements of the matrix $D_{\mathcal{I}}$. We set $r_i = b$ and $s_j = v$ for all $i \in [v], j \in [b]$ and
\begin{equation*}
\setlength{\nulldelimiterspace}{0pt}
c_{ij}=\left\lbrace\begin{IEEEeqnarraybox}[\relax][c]{l's}
0, &if $A_\mathcal{I}(i,j) = 0$,\\
\infty, &if $A_\mathcal{I}(i,j) = 1$.
\end{IEEEeqnarraybox}\right.
\end{equation*}

By this choice the condition in inequality \eqref{eq:thm_matrix} is trivially satisfied whenever $I,J$ are chosen such that there is a point in $I$ which is incident to some block in $J$, i.e., there exist $i \in I, j \in J$ such that $A_\mathcal{I}(i,j)=1$. Hence we restrict our attention to choices of $I$ and $J$ such that none of the points in $I$ are incident to any block in $J$. Under this restriction, the L.H.S. of inequality \eqref{eq:thm_matrix} is $0$ and the condition is equivalent to $(v-|I|)b \geq |J|v$. We will assume that
\begin{IEEEeqnarray}{C}\label{eq:thm_cap_condition}
  \exists I \subseteq [v], J \subseteq [b] ~\text{such that} \\
  A_\mathcal{I}(i,j)=0 ~\forall i \in I, j \in J, ~\text{and}~ (v-|I|)b < |J|v, \IEEEnonumber
\end{IEEEeqnarray}
and show that it leads to a contradiction for each of the three incidence structures considered.

If $\mathcal{I}$ corresponds to a $d$-regular simple graph, then $b=dv/2$. Consider the point-block incidence matrix $A_{\mathcal{I}}$, which is a $(0,1)$-matrix of size $v \times b$. For the chosen $I$ in eq. \eqref{eq:thm_cap_condition}, we look at the submatrix $A_{\mathcal{I}}[I, [b]]$ of size $|I|\times b$ that consists of the rows of $A_{\mathcal{I}}$ indexed by the points in $I$ and all the columns. Let $l_1$ be the number of columns with a single $1$ in $A_{\mathcal{I}}[I,[b]]$ and $l_2$ be the number of columns with two $1$s in $A_{\mathcal{I}}[I,[b]]$. By counting the total number of $1$s in $A_{\mathcal{I}}[I,[b]]$ in two ways, we get that
\begin{equation*}
  d|I| = l_1 + 2l_2 \leq 2(l_1+l_2) \implies l_1+l_2 \geq \frac{d|I|}{2}.
\end{equation*}
Since the number of edges incident to at least one point in $I$ is $l_1+l_2$, any subset $J$ of the edges that has no incidence with any point in $I$ satisfies $|J|\leq b-d|I|/2$. Using these in eq. \eqref{eq:thm_cap_condition} we get that
\begin{equation*}
(v-|I|)b < |J|v \implies (v-|I|)\frac{dv}{2} < \left( \frac{dv}{2} - \frac{d|I|}{2}\right)v,
\end{equation*} which is a contradiction.

Now suppose that $\mathcal{I}$ corresponds to a biregular bipartite graph, with $L$ vertices having degree $d_L$ in the left part and $R$ vertices having degree $d_R$ in the right part. Then $b = Ld_L = Rd_R$. Consider a subset $I_L \cup I_R$ of its vertices. Let $E_L$ (resp. $E_R$) be the set of edges which are incident to some vertex in $I_L$ (resp. $I_R$) but not incident to any vertex in $I_R$ (resp. $I_L$). The number of edges that are not incident to any vertex in $I_L \cup I_R$ is equal to $(L-|I_L|)d_L - |E_R| = (R-|I_R|)d_R - |E_L|$. Suppose there is a choice of $I$ in eq. \eqref{eq:thm_cap_condition} is such that $I=I_L \cup I_R$ for some $I_L, I_R$. Then we have that
\begin{IEEEeqnarray*}{L}
 (v-|I|)b < |J|v, \\
 \!\implies \!\! (L+R - (|I_L|+|I_R|))\frac{Ld_L + Rd_R}{2}\\ \hfill < \frac{(L-|I_L|)d_L-E_R + (R-|I_R|)d_R-|E_L|}{2}(L+R),\\
 \!\implies \!\! \frac{|I_l|d_L + |I_R|d_R + |E_L| + |E_R|}{Ld_L + Rd_R} < \frac{|I_L|+|I_R|}{L+R},\\%  \frac{(L-|I_L|)d_L - |E_R| +(R-|I_R|)d_R - |E_L|}{Ld_L + Rd_R} = 1 - \frac{|I_l|d_L + |I_R|d_R + |E_L| + |E_R|}{Ld_L + Rd_R},\\
%& \implies (Ld_L + Rd_R)(|I_L|+|I_R|) - (L+R)(I_Ld_L+I_Rd_R)  > (L+R)(|E_L|+|E_R|),\\
 \!\implies \!\! (L\!+\!R)(|E_L|\!+\!|E_R|) < (L\!-\!R)|I_L|d_L \! +\! (R\!-\!L)|I_R|d_R,\\
 \!\implies \!\! (L+R)(|E_L|+|E_R|) < (L-R)(|E_L|-|E_R|).
\end{IEEEeqnarray*}
If $L>R$ or $|E_L|>|E_R|$, then we have a contradiction. That leaves the case when $L < R$ and $|E_L| < |E_R|$, which implies $(L+R)(|E_L|+|E_R|) < (R-L)(|E_R|-|E_L|)$ and that is also a contradiction.

Next, consider a $t$-$(v,k,1)$ design with $b$ blocks such that repetition degree of each point is $\rho$ and we have that $bk=v\rho$. With the $I$ of eq. \eqref{eq:thm_cap_condition}, we employ a similar procedure as for the case of the $d$-regular graph. We choose the submatrix $A_{\mathcal{I}}[I,[b]]$ of size $|I|\times b$ that corresponds to the rows indexed by the points in $I$ and let $l_i, \forall i \in [k]$ denote the number of columns with exactly $i$ $1$s in $A_{\mathcal{I}}[I,[b]]$. We count the total number of $1$s in $A_{\mathcal{I}}[I,[b]]$ in two ways, yielding
\begin{IEEEeqnarray*}{C}
  \rho|I| = l_1 + 2l_2 + \cdots + (k-1)l_{k-1} + kl_k \leq k\sum_{i=1}^kl_i,\\
  \implies \sum_{i=1}^{k}l_i \geq \frac{\rho|I|}{k} = \frac{b|I|}{v}.
\end{IEEEeqnarray*}
The number of blocks that are incident to at least one point in $I$ is equal to $\sum_{i=1}^{k}l_i$. Hence any subset $J$ of blocks that has no incidence with any point in $I$ satisfies $|J|\leq b-|I|b/v$. Using this in eq. \eqref{eq:thm_cap_condition} we get that
\begin{equation*}
(v-|I|)b < |J|v \implies (v-|I|)b < \left(b-\frac{|I|b}{v}\right)v,
\end{equation*} which is a contradiction.

If $\mathcal{I}=(\calP,\mathcal{B})$ is the higher incidence structure obtained from a $t$-$(n,t+1,\lambda)$ design as described in corollary \ref{cor:tdesign_higher_inc}, then we have that $|\calP|=\binom{n}{t}$ and $|\mathcal{B}|=\frac{\lambda}{t+1}\binom{n}{t}$. By definition of $t$ for the original design, we have that each of the points in $\calP$ are incident to exactly $\lambda$ blocks. Also, each block in $\mathcal{B}$ consists of $\binom{t+1}{t}=t+1$ points. For the submatrix $A_{\mathcal{I}}[I,[b]]$ whose rows correspond to the points in $I$ from Condition $\ref{eq:thm_cap_condition}$, we let $l_i, \forall i \in [t+1]$ denote the number of columns that have exactly $i$ $1$s in them. By counting the total number of $1$s in $A_{\mathcal{I}}[I,[b]]$ in two ways we get that
\begin{equation*}
  \lambda |I| = \sum_{i=1}^{t+1}il_i \leq (t+1)\sum_{i=1}^{t+1}l_i \implies \sum_{i=1}^{t+1}l_i \geq \frac{\lambda |I|}{t+1}.
\end{equation*}
The total number of blocks incident to at least one point in $I$ is $\sum_{i=1}^{t+1}l_i$. Then the number of blocks $|J|$ that are not incident to any point in $I$ satisfy $|J|\leq |\mathcal{B}|-|I|\lambda/(t+1)$. Using these we get that
\begin{IEEEeqnarray*}{C}
(v-|I|)b < |J|v,\\
 \implies \left[ \binom{n}{t}-|I|\right]\frac{\lambda}{t+1}\binom{n}{t} < \frac{\lambda}{t+1}\left[\binom{n}{t}-|I|\right]\binom{n}{t},
\end{IEEEeqnarray*} which is a contradiction. Thus in all the three kinds of incidence structures considered, we have shown that they admit the existence of the associated matrix $D_\mathcal{I}$ under the stated qualifying conditions. This enables us to apply theorem \ref{thm:lb} and obtain a lower bound on the computation capacity of these sum-networks.
\end{IEEEproof}
For an undirected graph $\mathcal{I}=(\calP,\calB)$ that is not regular, proposition \ref{thm:capacity} is not applicable. Theorem \ref{thm:lb} describes a sufficient condition for the existence of a linear network code that achieves the upper bound on the computation capacity of normal sum-networks constructed from undirected graphs that are not necessarily regular. The upper bound on the capacity of the transposed sum-network constructed using the incidence matrix $A_\mathcal{I}^T$ however can be different from $\frac{|\calB|}{|\calB|+|\calP|}$ depending on the finite field $\mathcal{F}$ ({\it cf.} corollary \ref{cor:undirected_graph}) and theorem \ref{thm:lb} needs to be modified to be applicable in that case. The following example illustrates this.
\begin{example}\label{eg:irr-graph_transpose}
  Consider the transposed sum-network for the irregular graph $G$ described in Example \ref{eg:G}. Corollary \ref{cor:undirected_graph} gives an upper bound of $4/6$ on the computation capacity when $\mathcal{F}=GF(2)$, as for that case $\calP' = \{2,4\}$ and $\calB' = \{A,B,C,D\}$.
  We show the submatrix $A_G^T[\calB', \calP']$ in the equation below and also an associated matrix $D_G$ whose support is the same as that of $A_G^T[\calB', \calP']$ and whose row-sum $= 6-4 = 2$ and column-sum $= 4$.
  %The relevant submatrix $A_G^T[\calB', \calP']$ satisfies the condition in Proposition \ref{claim:irr-graph_lb} as there exists a matrix $D_G$ with the required row and column sums demonstrated below.
   The rows and columns are arranged in increasing alphabetical and numeric order.
  \begin{equation*}
    A_G^T[\calB', \calP'] = \begin{bmatrix}
                              1 & 0 \\
                              1 & 0 \\
                              0 & 1 \\
                              0 & 1
                            \end{bmatrix},
    D_G = \begin{bmatrix}
            2 & 0 \\
            2 & 0 \\
            0 & 2 \\
            0 & 2
          \end{bmatrix}.
  \end{equation*} Using $D_G$ we can construct a rate-$4/6$ linear network code, shown in Table \ref{tab:G-transpose_code}, that achieves the computation capacity for $\mathcal{F}=GF(2)$ of the transposed sum-network constructed using the irregular graph $G$ shown in Figure \ref{fig:G}. In particular, terminals $t_1, t_3$ don't need any information other than the partial sums obtained over their respective bottleneck edges to compute the sum. Terminals $t_2, t_4$ need the value $X_2, X_4$ respectively, and that is transmitted in a piecewise fashion according to the matrix $D_G$ over the bottleneck edges.
\begin{table*}
\caption{The function values transmitted across the bottleneck edges of the transposed sum-network corresponding to the graph shown in Figure \ref{fig:G} for a rate-$4/6$ network over $GF(2)$. Each message $X_2,X_4$ is a vector with $4$ components, and $\phi_A(X), \phi_B(X), \phi_C(X), \phi_D(X), \phi_E(X)$ are vectors with $6$ components each. The number inside square brackets adjoining a vector indicates a particular component of the vector. A dash indicates that the value transmitted on that component is not used in decoding by any terminal.}
\label{tab:G-transpose_code}
\centering
\begin{tabular}{cccccc}
\hline\hline
Component & $\phi_A(X)$ & $\phi_B(X)$ & $\phi_C(X)$ & $\phi_D(X)$ & $\phi_E(X)$\\ [0.5ex]
\hline
$1$ to $4$ & $X_1+X_2+X_A$ & $X_2+X_3+X_B$ & $X_3+X_4+X_C$ & $X_1+X_4+X_D$ & $X_1+X_3+X_E$\\
$5$ & $X_2[1]$ & $X_2[3]$ & $X_4[1]$ & $X_4[3]$ & --\\
$6$ & $X_2[2]$ & $X_2[4]$ & $X_4[2]$ & $X_4[4]$ & --\\
[1ex]
\hline
\end{tabular}
\end{table*}
\vspace{2mm}
\end{example}

For an undirected graph $\mathcal{I}=(\calP,\calB)$ that is not regular, let $\calP', \calB'$ be the set of points and edges as chosen in the statement of corollary \ref{cor:undirected_graph}.
We describe a condition on the submatrix $A^T_\mathcal{I}[\calB',\calP']$ which consists of the rows and columns of $A_\mathcal{I}^T$ corresponding to the blocks and points in the sets $\calB',\calP'$ respectively. This condition allows us to construct a capacity-achieving linear network code for the transposed sum-network.
%\aditya{read this later}
\begin{proposition}\label{claim:irr-graph_lb}
For an undirected graph $\mathcal{I}=(\calP,\calB)$, let $|\calP'|=v', |\calB'|=b'$, where $\calP',\calB'$ are subsets of points and blocks as defined in corollary \ref{cor:undirected_graph} and let $A^T_\mathcal{I}[\calB',\calP'](i,j)$ indicate an element of the submatrix for indices $i \in [b'], j \in [v']$. Suppose there is a matrix $D_\mathcal{I}$ of dimension $b'\times v'$ such that
\begin{IEEEeqnarray*}{Rl}
D_\mathcal{I}(i,j)=& 0, ~\text{if}~ A^T_\mathcal{I}[\calB',\calP'](i,j)=0,\\
\sum_{i=1}^{b'}D_\mathcal{I}(i,j)=& b', ~\text{for all}~ j \in [v'], ~\text{and}\\
\sum_{j=1}^{v'}D_\mathcal{I}(i,j)=& v', ~\text{for all}~ i \in [b'].
\end{IEEEeqnarray*}
Then there is linear network code of rate $\frac{b'}{b'+v'}$ that allows each terminal in the transposed sum-network constructed using $\mathcal{I}$ to compute the required sum.
\end{proposition}
\begin{IEEEproof}
We describe a rate-$b'/(b'+v')$ network code that enables each terminal to compute the sum. Then by corollary \ref{cor:undirected_graph} we know that this is a capacity-achieving code. Since this is a transposed sum-network, the bottleneck edges in the sum-network correspond to the blocks in the undirected graph $\mathcal{I}$. The first $b'$ components transmitted over each bottleneck is obtained by the following equation.
\begin{equation*}
  \phi_i(X)[1:b']= X_{B_i} + \sum_{j:p_j \in B_i}X_{p_j}, ~\text{for all}~ B_i \in \calB.
\end{equation*}
We show that this partial sum satisfies all the terminals in the set $\{t_{B_i}: B_i \in \calB\}\cup \{t_{p_j}:p_j \notin \calP'\}$. Terminals in $\{t_{B_i}:B_i \in \calB\}$ can recover the sum as all messages not present in the partial sum are available to $t_{B_i}$ through direct edges. For terminals in the set $\{t_{p}:p \notin \calP'\}$, they carry out the following operation as a part of their decoding procedure.
\begin{IEEEeqnarray}{L}
\sum_{i:B_i \in \langle p\rangle}\phi_i(X)[1:b']=\sum_{i:B_i \in \langle p\rangle} \left( X_{B_i} + \sum_{j:p_j \in B_i} X_{p_j}\right) \IEEEeqnarraynumspace \label{eq:t_p-decode}\\
=\sum_{i:B_i \in \langle p \rangle}X_{B_i}+\sum_{j:\{p,p_j\}\in \calB}\bm{p}\bm{p}_j^T X_{p_j} + \deg(p)X_p. \IEEEeqnarraynumspace \label{eq:t_p-decode-sum}
\end{IEEEeqnarray}
For $p_j \neq p$, we have that $\bm{p}\bm{p}_j^T=1$ if $\{p,p_j\}\in \calB$. Also by condition on the points that are not in $\calP'$, we have that $\deg(p) \equiv 1 \pmod{\ch(\mathcal{F})}$, and hence all the coefficients in the above partial sum are $1$. The messages in the set $\{X_{B}:B \notin \langle p\rangle\} \cup \{X_{p_j}:\{p_j,p\}\notin \calB\}$ are available to $t_{p}$ through direct edges and hence it can recover the sum.

The remaining $v'$ components available on the bottleneck edges $\{e_i : B_i \in \calB'\}$ are used to transmit information that enable the terminals in the set $\{t_{p}:p \in \calP'\}$ to compute the sum. Specifically, we construct a flow on a bipartite graph whose one part corresponds to the points in $\calP'$ and the other part corresponds to the blocks in $\calB'$, with incidence being determined by the submatrix $A^T_{\mathcal{I}}[\calB',\calP']$. Since there exists a matrix $D_\mathcal{I}$ with specified row and column sums, we can use it to construct a flow on the bipartite graph such that the messages in the set $\{X_{p_i}: p_i \in \calP'\}$ are transmitted in a piecewise fashion over the bottleneck edges $\{e_j:B_j \in \calB'\}$ in a manner similar to the proof of theorem \ref{thm:lb}. Arguing in the same way, one can show that the network code based on the flow solution allows each $t_p ~\forall p \in \calP'$ to obtain the value of $X_p$ from the information transmitted over the bottleneck edges in the set $\{e_i: B_i \in \langle p \rangle\}$. Terminal $t_p$ computes the sum in eq. \eqref{eq:t_p-decode} as a part of its decoding procedure. Since $\deg(p) \not\equiv 1 \pmod{\ch(\mathcal{F})}$, every term in the RHS of eq. \eqref{eq:t_p-decode-sum} except $X_p$ has its coefficient as $1$. But since $t_p$ knows the value of $X_p$ it can subtract a multiple of it and recover the relevant partial sum.
% Since $A_{\mathcal{I}}[\calP',\calB']A^T_{\mathcal{I}}[\calP',\calB'] - \left(A_{\mathcal{I}}[\calP',\calB']A^T_{\mathcal{I}}[\calB',\calP']\right)_\#$ has all of its off-diagonal terms as zero and its diagonal terms as non-zero elements of $\mathcal{F}$, the information transmitted over the remaining $v'$ components as determined by the value of the flow allows each terminal $t_{p}, p \in \calP'$ to compute the value of $\sum_{B \in \langle p \rangle}X_B + \sum_{p': \{p',p\}\in \calB}X_{p'}$.
 The messages not present in this partial sum are available to $t_p$ through direct edges and hence it can also compute the value of the sum.
\end{IEEEproof}

%\aditya{I think this example should come first and the full proof later}

Proposition \ref{thm:capacity} describes families of incidence structures for which the sum-networks constructed admit capacity-achieving linear network codes. The upper bound on the computation capacity of these sum-networks is obtained from Corollaries \ref{cor:undirected_graph}, \ref{cor:bibd} and \ref{cor:tdesign_higher_inc}. We now describe a rate-$1$ linear network code for the sum-networks when their corresponding incidence structures do not satisfy the qualifying conditions for the upper bounds. By theorem \ref{thm:ub_all}, the computation capacity of any sum-network obtained using the SUM-NET-CONS algorithm is at most $1$.
\begin{proposition}\label{cor:scalar_nw_code}
  For an incidence structure $\mathcal{I}=(\calP,\calB)$ and a finite field $\mathcal{F}$, there exists a rate-$1$ linear network code that satisfies the following listed sum-networks. If
  \begin{itemize}
    \item $\mathcal{I}$ is a $2$-$(v,k,1)$ design:
     \begin{itemize}
       \item the normal sum-network with $\ch(\mathcal{F})\mid k-1$,
       \item the transpose sum-network with $\ch(\mathcal{F})\mid \frac{v-k}{k-1}$,
     \end{itemize}
    \item $\mathcal{I}$ is a $t$-$(v,t+1,\lambda)$ design:
    \begin{itemize}
      \item the normal sum-network obtained using the higher incidence matrix with $\ch(\mathcal{F})\mid t$,
      \item the transpose sum-network obtained using the higher incidence matrix with $\ch(\mathcal{F})\mid \lambda-1$.
    \end{itemize}
  \end{itemize}
\end{proposition}
\begin{IEEEproof}
Suppose we construct a sum-network using the SUM-NET-CONS algorithm on a $(0,1)$-matrix $A$ of dimension $r \times c$. If $A^TA=(A^TA)_\#$, the following rate-$1$ linear network code
\begin{equation*}
  \phi_i(X) = X_{p_i} + \sum_{j:B_j \in \langle p_i\rangle}X_{B_j}, ~\forall~ i \in [r],
\end{equation*}satisfies every terminal in the sum-network in the following manner.
A terminal $t_{p_i}, ~\forall i \in [r]$ receives all the messages not present in the partial sum transmitted along $e_i$ through direct edges, and hence it can compute the sum. A terminal $t_{B}, ~\forall B \in \mathcal{B}$ can carry out the following operation.
\begin{IEEEeqnarray*}{Rl}
  \sum_{i:p_i \in B_j}\phi_i(X) &= \sum_{p_i \in B} X_{p_i} + \sum_{p_i \in B}\sum_{B_j \in \langle p_i\rangle}X_{B_j}\\
   &= \sum_{p_i \in B} X_{p_i} + \sum_{l:B_l \in \langle B_j\rangle}\bm{B}_l^T\bm{B}_j X_{B_l}.
\end{IEEEeqnarray*} %where $\bm{B}_l$ for any $l \in [c]$ indicates the $l$-th column of $A$.
Since $A^TA=(A^TA)_\#$, all the coefficients in the above sum are $1$ and $\sum_{i:p_i \in B_j}\phi_i(X)$ is equal to the sum of all the messages in the set $\{X_{p_i}: p_i \in B_j\}\cup \{X_B: B \in \langle B_j\rangle\}$. All the messages that are not present in this set are available to $t_{B_j}$ through direct edges.

Such a rate-$1$ linear network code gives us our proposition in the following manner.
Let $A_\mathcal{I}$ be the $v \times \frac{v-1}{k-1}$ incidence matrix for a $2$-$(v,k,1)$ design and let $A_\mathcal{I}'$ be the higher incidence matrix as defined in corollary \ref{cor:bibd} for a $t$-$(v,t+1,\lambda)$ design with $\lambda \neq 1$. Then, we have (from proofs of Corollaries \ref{cor:bibd}, \ref{cor:tdesign_higher_inc})
\begin{IEEEeqnarray*}{Rl}
A_\mathcal{I}^TA_\mathcal{I}-(A_\mathcal{I}^TA_\mathcal{I})_\# =& (k-1)I,\\
A_\mathcal{I}A_\mathcal{I}^T-(A_\mathcal{I}A_\mathcal{I}^T)_\# =& \frac{v-k}{k-1}I,\\
A_\mathcal{I}^{'T}A'_\mathcal{I}-(A_\mathcal{I}^{'T}A'_\mathcal{I})_\# =& tI,\\
A'_\mathcal{I}A_\mathcal{I}^{'T}-(A'_\mathcal{I}A_\mathcal{I}^{'T})_\# =& (\lambda -1)I.
\end{IEEEeqnarray*}
Thus, whenever any of the above matrices is a zero matrix, we have a scalar linear network code that achieves the computation capacity of the associated sum-network.
\end{IEEEproof}

\section{Discussion and comparison with prior work}\label{sec:discussion}
The discussion in Sections \ref{sec:upper_bd} and \ref{sec:lower_bd} establishes the computation capacity for sum-networks derived from several classes of incidence structures. We now discuss the broader implications of these results by appealing to existence results for these incidence structures. BIBDs have been the subject of much investigation in the literature on combinatorial designs. In particular, the following two theorems are well-known.
\begin{theorem}\label{thm:steiner_triple_exist}
\cite[Theorem 6.17]{Stinson} There exists a $(v,3,1)$-BIBD (also known as a Steiner triple system) if and only if $v \equiv 1,3 \pmod 6; v\geq 7$.
\end{theorem}
%\aditya{check following thm}
\begin{theorem}\label{thm:v-4-1_bibd}
  \cite[Theorem 7.31]{Stinson} There exists a $(v,4,1)$-BIBD if and only if $v \equiv 1,4 \pmod{12}; v \geq 13$.
\end{theorem}
In particular, these results show that there are an infinite family of Steiner triple systems and BIBDs with block size $4$ and $\lambda=1$.
Since $k=3$ for any Steiner triple system, we can demonstrate the existence of sum-networks whose computation capacity is greatly affected by the choice of the finite field $\mathcal{F}$ used for communication.
%Applying our results in Sections REF, we obtain the following result.
\begin{proposition}\label{prop:steiner_system}
Consider the normal sum-network constructed using a $2$-$(v,3,1)$ design. If $\ch(\mathcal{F})=2$, then the computation capacity of the sum-network is $1$. For odd $\ch(\mathcal{F})$, the computation capacity is $\frac{6}{5+v}$. For the normal sum-network constructed using a $(v,4,1)$-BIBD, the computation capacity is $1$ if $\ch(\mathcal{F})=3$ and $\frac{12}{11+v}$ otherwise.
\end{proposition}
\begin{IEEEproof}
The number of blocks in a $2$-$(v,3,1)$ design is equal to $v(v-1)/6$. From corollary \ref{cor:bibd}, if $\ch(\mathcal{F})$ is odd, then the computation capacity of the sum-network constructed using a Steiner triple system is at most $\frac{v}{v+v(v-1)/6}=\frac{6}{5+v}$. Moreover by proposition \ref{thm:capacity}, we can construct a linear network code with rate equal to the upper bound. On the other hand, if $\ch(\mathcal{F})=2$, then the computation capacity of the same sum-network is $1$ by proposition \ref{cor:scalar_nw_code}.

The number of blocks in a $2$-$(v,4,1)$ design is $v(v-1)/12$. We can recover the result for the computation capacity of a normal sum-network constructed using it in a manner similar to the previous case.
\end{IEEEproof}
Thus, this result shows that for the {\it same} network, computing the sum over even characteristic has capacity $1$, while the capacity goes to zero as $O(1/v)$ for odd characteristic. Moreover, this dichotomy is not unique to the prime number $2$. Similar results hold for sum-networks derived from higher incidence structures (\textit{cf.} corollary \ref{cor:tdesign_higher_inc}).
\begin{theorem}\label{t_design_exist}
\cite{teirlinck87} For two integers $t,v$ such that $v \geq t+1 > 0$ and $v \equiv t \pmod {(t+1)!^{2t+1}}$, a $t$-$(v,t+1,(t+1)!^{2t+1})$ design with no repeated blocks exists.
\end{theorem}
The number of blocks in a $t$-$(v,t+1,(t+1)!^{2t+1})$ design can be evaluated to be $\binom{v}{t}\frac{(t+1)!^{2t+1}}{t+1}$. We consider the normal sum-network obtained using the higher incidence matrix of this $t$-design. If $\ch(\mathcal{F})\nmid t$, then by corollary \ref{cor:tdesign_higher_inc} and proposition \ref{thm:capacity}, we have that the computation capacity of this sum-network is
\begin{equation*}
  \frac{\binom{v}{t}}{\binom{v}{t}+\binom{v}{t}\frac{(t+1)!^{2t+1}}{t+1}}=\frac{1}{1+t!^2(t+1)!^{2t-1}}.
\end{equation*}
On the other hand, if $\ch(\mathcal{F})$ is a divisor of $t$, then by theorem \ref{thm:ub_all} and proposition \ref{cor:scalar_nw_code} we have that the computation capacity of the normal sum-network constructed using the higher incidence matrix is $1$. Thus for the same sum-network, computing the sum over a field whose characteristic divides the parameter $t$ can be done at rate $=1$. However, if the field characteristic does not divide $t$, zero-error computation of the sum can only be done at a rate which goes to zero as $O\left(\left(\frac{t}{e}\right)^{-t^2}\right)$.

Theorem \ref{thm:steiner_triple_exist} describes an infinite family of BIBDs with $k=3$ and $\lambda =1$. There are further existence results for BIBDs with $\lambda =1$ and $k\neq 3$. In particular, for $\lambda =1, k \leq 9$ there exist BIBDs with value of $v$ as given in Table 3.3 in \cite[Section II.3.1]{handbook_colbourn06}. As an example, if $k = 5$, then there exists a $2$-$(v,5,1)$ design whenever $v \equiv 1,5 \pmod 20$. For any choice of a BIBD from this infinite family, we can construct a corresponding normal sum-network, whose computation capacity for a particular finite field can be found using corollary \ref{cor:bibd} and proposition \ref{thm:capacity}.
%Other existence results for incidence structures considered above with different values for their parameters are also known , \cite[Section II.4.3]{handbook_colbourn06}. \aditya{quite vague. We need to be more precise here}
Even though theorem \ref{t_design_exist} states the existence of $t$-designs for $v,t$ that satisfy the qualifying conditions, explicit constructions of such $t$-designs with $t\geq 6$ are very rare.

For a transposed sum-network obtained from an undirected graph that is not regular, the computation capacity can show a more involved dependence on the finite field alphabet as the following example demonstrates.
\begin{example}\label{eg:irr_graph}
  \begin{figure}
    \centering
    \includegraphics[scale = 0.65]{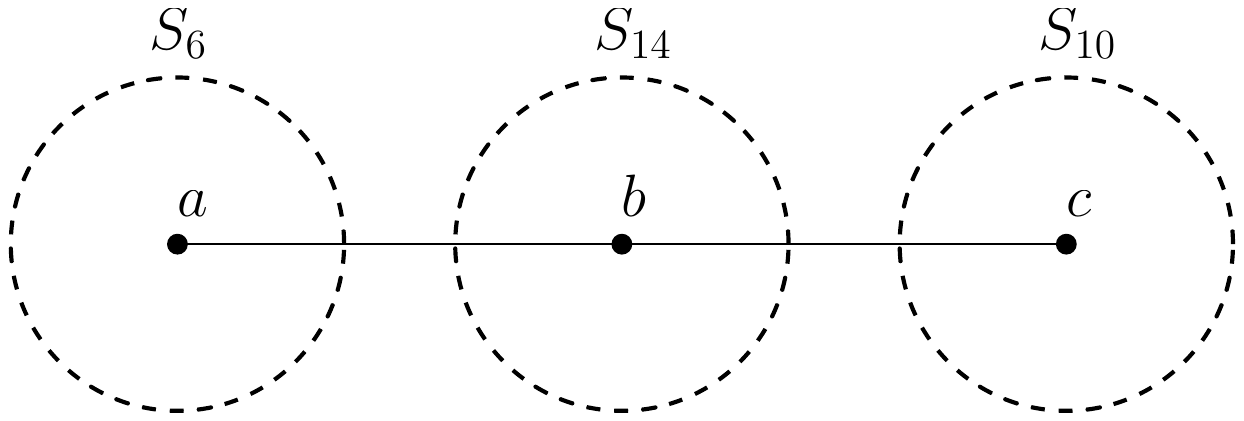}
    \caption{The schematic shown represents an undirected graph with three components: $S_{6}$, $S_{14}$ and $S_{10}$. $S_{t}$ denotes the star graph on $t+1$ vertices, with only one vertex having degree $t$ while the rest have degree $1$. The vertices with the maximum degree in the three star graphs are $a,b,c$ respectively. In addition, $a$ is connected to $b$ and $b$ is connected to $c$, such that $\deg(a)=7, \deg(b)=16, \deg(c)=11$.}\label{fig:irregular_eg}
  \end{figure}
  Consider the transposed sum-network obtained by applying the \sumnetalg algorithm on the undirected graph $\mathcal{I}$ shown in Figure \ref{fig:irregular_eg}. Corollary \ref{cor:undirected_graph} gives us an upper bound on the computation capacity of the transposed sum-network based on the finite field alphabet $\mathcal{F}$. The upper bound for three different choices of $\mathcal{F}$ is as follows.
  \begin{itemize}
    \item $\mathcal{F}=GF(2)$: Then $\mathcal{P}'=\{b\}$, so the upper bound is $16/(16+1)=16/17$.
    \item $\mathcal{F}=GF(3)$: Then $\mathcal{P}'=\{c\}$, so the upper bound is $11/(11+1)=11/12$.
    \item $\mathcal{F}=GF(5)$: Then $\mathcal{P}'=\{a\}$, so the upper bound is $7/(7+1)=7/8$.
  \end{itemize}
We use proposition \ref{claim:irr-graph_lb} to check if we can construct a linear network code in each case that has the same rate as the respective upper bound. To do that, we focus on the appropriate submatrix of $A_\mathcal{I}$ for each case and see if it satisfies the required condition on row and column sums. The rows of $A_{\mathcal{I}}$ corresponding to the vertices $a, b, c$ (in order) are shown below.
  \begin{equation*}
    \begin{bmatrix}
      \bm{1}_6 & 1 & 0 & \cdots & 0 \\
      \bm{0}_6 & 1 & \bm{1}_{14} & 1 & \bm{0}_{10} \\
      0 & \cdots & 0 & 1 & \bm{1}_{10}
    \end{bmatrix},
  \end{equation*} where $\bm{1}, \bm{0}$ indicate all-one and all-zero row vectors of size specified by their subscripts. Using this, one can verify that the appropriate submatrix for each of the three choices of $\mathcal{F}$ satisfies the conditions of proposition \ref{claim:irr-graph_lb} and hence we can construct a capacity-achieving linear network code in each case.
\vspace{2mm}
\end{example}

Thus, as the previous example demonstrates, the computation capacity of a particular sum-network need not take just one of two possible values, and can have a range of different values based on the finite field chosen. We can generalize the example to obtain sum-networks that have arbitrary different possible values for their computation capacity.

Our constructed sum-networks have a unit maximum flow between any source and any terminal. We can modify our construction so that each edge in the network has a capacity of $\alpha > 1$. Specifically, the following result can be shown.
\begin{proposition}\label{prop:multi_cap_edge}
Let $\mathcal{N}$ denote the sum-network obtained by applying the \sumnetalg algorithm on a matrix $A$ of dimension $r \times c$. For an integer $\alpha >1$, let $\mathcal{N}_\alpha$ denote the sum-network obtained by modifying the \sumnetalg algorithm such that $\mathcal{N}_\alpha$ has the same structure as $\mathcal{N}$ but each edge $e_\alpha$ in $\mathcal{N}_\alpha$ has $\capacity(e_\alpha)=\alpha >1$. %each edge added in the procedure has capacity $\alpha$ instead of $1$.
Then, if $A$ satisfies the qualifying conditions in Theorems \ref{thm:ub} and \ref{thm:lb}, the computation capacity of $\mathcal{N}_\alpha$ is $\frac{\alpha r}{r+c}$.
\end{proposition}
\begin{IEEEproof}
Since $A$ satisfies the conditions in theorem \ref{thm:lb}, there exists a $(m,n)$ vector linear network code with $m=r, n=r+c$. For every unit-capacity edge in $\mathcal{N}$, we have $\alpha$ unit-capacity edges between the same tail and head in $\mathcal{N}_\alpha$. At the tail of every edge in $\mathcal{N}_\alpha$, we can apply the same network code except now we have $\alpha$ distinct edges on which we can transmit the encoded value. Thus we need transmit only $\frac{r+c}{\alpha}$ symbols on each of those edges. If $\frac{r+c}{\alpha}$ is not an integer, one can appropriately multiply both $m,n$ with a constant. This modified network code has rate $=\frac{\alpha r}{r+c}$. Since $A$ also satisfies the conditions in theorem \ref{thm:ub}, we have that an upper bound on the computation capacity of $\mathcal{N}$ is $r/(r+c)$. Applying the same argument on $\mathcal{N}_\alpha$, we get that an upper bound on the computation capacity of $\mathcal{N}_\alpha$ is $\frac{\alpha r}{r+c}$. This matches the rate of the modified vector linear network code described above.
\end{IEEEproof}

This result can be interpreted as follows. Consider the class of sum-networks where the maximum flow between any source-terminal pair is at least $\alpha$. Our results indicate, that for any $\alpha$, we can always demonstrate the existence of a sum-network, where the computation capacity is strictly smaller than $1$. Once again, this indicates the crucial role of the network topology in function computation.

\subsection{Comparison with prior work}\label{subsec:comparison}
The work of Rai and Das \cite{raiD13} is closest in spirit to our work. In \cite{raiD13}, the authors gave a construction procedure to obtain a sum-network with computation capacity equal to $p/q$, where $p,q$ are any two co-prime natural numbers. The procedure involved first constructing a sum-network whose capacity was $1/q$. Each edge in this sum-network had unit-capacity. By inflating the capacity of each edge in the sum-network to $p > 1$, the modified sum-network was shown to have computation capacity as $p/q$.

Our work is a significant generalization of their work. In particular, their sum-network with capacity $1/q$ can be obtained by applying the \sumnetalg algorithm to the incidence matrix of a complete graph on $2q-1$ vertices \cite{tripathyR14}. We provide a systematic procedure for constructing these sum-networks for much larger classes of incidence structures.

In \cite{raiD13}, the authors also posed the question if \textit{smaller} sum-networks (with lesser sources and terminals) with capacity as $p/q$ existed. Using the procedure described in this paper, we can answer that question in the affirmative. %The following instance illustrates that.
\begin{example}\label{eg:smaller}
The normal sum-network for the undirected graph in Figure \ref{fig:G} has computation capacity $=4/9$ and has nine sources and terminals. To obtain a sum-network with the same computation capacity using the method described in \cite{raiD13} would involve constructing the normal sum-network for a complete graph on $17$ vertices, and such a sum-network would have $153$ source nodes and terminal nodes each.
\vspace{2mm}
\end{example}

In \cite{ramamoorthyL13}, it was shown by a counter-example that for the class of sum-networks with $|S|=|T|=3$, a maximum flow of $1$ between each source-terminal pair was not enough to guarantee solvability (i.e., no network code of rate $1$ exists for the counterexample). It can be observed that their counter-example is the sum-network shown in Figure \ref{fig:K2normal}. Our characterization of computation capacity for a family of sum-networks provides significantly more general impossibility results in a similar vein.
In particular, note that for the $\alpha$-capacity edge version of a sum-network, the maximum flow between any source-terminal pair is at least $\alpha$. Then suppose we consider the class of sum-networks with $|S|=|T|=x=\beta(\beta+1)/2$ for some $\beta \in \mathbb{N}$. Consider a complete graph $K_\beta=(V,E)$ on $\beta$ vertices; then $|V|+|E|=x$. Consider the sum-network obtained by applying the procedure on $K_\beta$, with each edge added having capacity as $\alpha$. Then the computation capacity of this sum-network is $\alpha \beta/x$, which is less than $1$ if $\alpha < (\beta+1)/2$. This implies that a max-flow of $(\beta+1)/2$ between each source-terminal pair is a necessary condition for ensuring all sum-networks with $|S|=|T|=x$ are solvable. When $x$ cannot be written as $\beta(\beta+1)/2$ for some $\beta$, a similar argument can be made by finding an undirected graph $G=(V,E)$ (whose incidence matrix $A_G$ satisfies the condition in theorem \ref{thm:lb}) such that $|V|$ is minimal and $|V|+|E|=x$.

\section{Conclusions and Future Work}\label{sec:conclusion}
%\aditya{needs some work}
Sum-networks are a large class of function computation problems over directed acyclic networks. The notion of computation capacity is central in function computation problems, and various counterexamples and problem instances have been used by different authors to obtain a better understanding of solvability and computation capacity of general networks. We provide an algorithm to systematically construct sum-network instances using combinatorial objects called incidence structures. We propose novel upper bounds on their computation capacity, and in most cases, give matching achievable schemes that leverage results on the existence of non-negative integer matrices with prescribed row and column sums. We demonstrate that the dependence of computation capacity on the underlying field characteristic can be rather strong.

There are several opportunities for future work. Our proposed linear network codes for the constructed sum-networks require the corresponding incidence structures to have a specific property. In particular, our techniques only work in the case when $A^TA- (A^TA)_\#$ is a diagonal matrix. It would be interesting to find capacity achieving network codes in cases when $A^TA- (A^TA)_\#$ is not diagonal. More generally, it would be interesting to obtain achievability schemes and upper bounds for sum-networks with more general topologies.

%\section*{Acknowledgment}
%
%
%The authors would like to thank...

\bibliographystyle{IEEEtran}
% argument is your BibTeX string definitions and bibliography database(s)
%\bibliography{IEEEabrv,../BibFiles/tip}
\bibliography{IEEEabrv,tip}
%
% <OR> manually copy in the resultant .bbl file
% set second argument of \begin to the number of references
% (used to reserve space for the reference number labels box)
%\begin{thebibliography}{1}
%
%\bibitem{IEEEhowto:kopka}
%H.~Kopka and P.~W. Daly, \emph{A Guide to \LaTeX}, 3rd~ed.\hskip 1em plus
%  0.5em minus 0.4em\relax Harlow, England: Addison-Wesley, 1999.
%
%\end{thebibliography}
\vspace{-0.5in}
\begin{IEEEbiographynophoto}{Ardhendu Tripathy}
(S'15) is a Ph.D. student in the Department of Electrical and Computer Engineering at Iowa State University. He obtained his B.Tech. degree in Electrical Engineering from the Indian Institute of Technology, Kanpur in 2012. His research interests are in the areas of information theory, machine learning and signal processing.
\end{IEEEbiographynophoto}
\begin{IEEEbiographynophoto}{Aditya Ramamoorthy}
(M'05) received the B.Tech. degree in electrical engineering from the Indian Institute of Technology, Delhi, in 1999, and the M.S. and Ph.D. degrees from the University of California, Los Angeles (UCLA), in 2002 and 2005, respectively. He was a systems engineer with Biomorphic VLSI Inc. until 2001. From 2005 to 2006, he was with the Data Storage Signal Processing Group of Marvell Semiconductor Inc. Since fall 2006, he has been with the Electrical and Computer Engineering Department at Iowa State University, Ames, IA 50011, USA. His research interests are in the areas of network information theory, channel coding and signal processing for bioinformatics and nanotechnology. Dr. Ramamoorthy served as an editor for the IEEE Transactions on Communications from 2011 -- 2015. He is currently serving as an associate editor for the IEEE Transactions on Information Theory. He is the recipient of the 2012 Early Career Engineering Faculty Research Award from Iowa State University, the 2012 NSF CAREER award, and the Harpole-Pentair professorship in 2009 and 2010.
\end{IEEEbiographynophoto}

% Can use something like this to put references on a page
% by themselves when using endfloat and the captionsoff option.
\ifCLASSOPTIONcaptionsoff
  \newpage
\fi

\end{document}